\documentclass[a4paper,fleqn,usenatbib]{mnras}

\usepackage[T1]{fontenc}
\usepackage{ae,aecompl}

\usepackage{graphicx}	
\usepackage{subfig}
\usepackage{amsmath}	
\usepackage{amssymb}	
\usepackage{siunitx}
\usepackage{textcomp}

\graphicspath{{Plots/}}

\DeclareMathOperator{\sign}{sgn}

\title[Dusty warped discs]{Radiative transfer modelling of parsec-scale dusty warped discs}

\author[H.~Jud et al.]
  {H.~Jud,$^{1}$\thanks{E-mail: hansueli.jud@outlook.com}
   M.~Schartmann,$^{1}$
   J.~Mould,$^{1}$
   L.~Burtscher$^{2}$
   and K.~R.~W.~Tristram$^{3}$
\\
$^{1}$Centre for Astrophysics and Supercomputing, Swinburne University
of Technology, P.~O.~Box 218, Hawthorn, Victoria 3122, Australia\\
$^{2}$Max-Planck-Institut f\"ur extraterrestrische Physik, Postfach 1312, Gie{\ss}enbachstr., 85741 Garching, Germany\\
$^{3}$European Southern Observatory, Alonso de C\'{o}rdova 3107, Vitacura, Casilla 19001, Santiago de Chile, Chile
}

\date{Accepted XXX. Received YYY; in original form ZZZ}

\pubyear{2016}

\begin{document}
\label{firstpage}
\pagerange{\pageref{firstpage}--\pageref{lastpage}}
\maketitle

\begin{abstract}
Warped discs have been found on (sub-)\,parsec 
scale in some nearby Seyfert nuclei, identified by their maser emission.
Using dust radiative transfer simulations we explore their observational signatures in the infrared in order to find out whether  
they can partly replace the molecular torus. Strong variations of the brightness distributions 
are found, depending on the orientation of the warp with respect to the line of sight. Whereas images at short wavelengths typically
show a disc-like and a point source component, the warp itself only becomes visible at far-infrared wavelengths.
A similar variety is visible in the shapes of the spectral energy distributions. 
Especially for close to edge-on views,
the models show silicate feature strengths ranging from deep absorption to strong emission for variations of the lines of sight towards the warp. 
To test the applicability of our model, 
we use the case of the Circinus galaxy, where infrared interferometry has revealed a highly elongated emission component matching 
a warped maser disc in orientation and size. Our model is for the first time able to present a physical explanation for the observed dust 
morphology as coming from the AGN heated dust.
As opposed to available torus models, a warped disc morphology produces a variety of 
silicate feature shapes for grazing lines of sight, close to an edge-on view. This could be an attractive alternative to a claimed change 
of the dust composition for the case of the nearby Seyfert~2 galaxy NGC~1068, which harbours a warped maser disc as well.
\end{abstract}






\section{Introduction and motivation}
\label{sec:introduction}

Due to their vicinity, Seyfert galaxies have been the prime target to investigate the physics of Active Galactic Nuclei (AGN). 
An integral part of our current understanding of AGN is a geometrically thick gas and dust structure -- typically 
referred to as the molecular torus. A toroidal shape enables the orientation-dependent obscuration \citep{Antonucci_93,Urry_95},
put forward to unify two observed classes of active galaxies: type~1 sources have direct lines of sight towards the 
centre and show both, narrow and broad emission lines and type~2 sources which show narrow emission lines only.
The torus is also the reservoir of gas available for accretion
onto a central supermassive black hole (SMBH) which is thought to power the central activity. A fraction of the emitted
UV/X-ray is reprocessed into the IR wavelength regime by the dust, visible in the so-called IR-bump in their spectral
energy distributions (SEDs).
The idea of such a geometrically thick dust structure was coined after broad emission lines have been found in the polarised 
light of type~2 sources \citep[e.~g.~][]{Miller_83,Antonucci_85}. 
This led to the interpretation that a type~1 nucleus is hidden inside type~2 sources by a dusty torus seen edge-on.
Radiation is scattered towards the 
observer by free electrons and tenuous dust in the opening cone of the torus.

First direct evidence for the geometrical distribution of the dust was found by \citet{Jaffe_04} with the help of the mid-infrared interferometric instrument \citep[MIDI,][]{Leinert_03} 
at the Very Large Telescope Interferometer (VLTI). 
One of the best-studied parsec-sized dust distribution is harboured by the Circinus galaxy, which is the 
closest \citep[4.2\,Mpc, 1~arcsec $\approx$ 20~pc,][]{Freeman_77} and the second brightest Seyfert galaxy in the MIR. 
High resolution mid-infrared interferometric observations have revealed a two-component morphology in the brightness distribution: 
(i) a larger scale, elongated component in direction of the ionisation cone (along PA$\approx$107$^{\circ}$) with a full-width at half-maximum (FWHM) size of 
roughly 0.8 $\times$ 1.9\,pc, which is responsible for 80\% of the total MIR flux and was interpreted either as the directly illuminated funnel walls of the dusty 
molecular torus or as part of the (filamentary/clumpy) outflow; (ii) a disc-like component with a FWHM size of roughly 0.2 $\times$ 1.1\,pc and elongated along PA$\approx$46$^{\circ}$ 
\citep{Tristram_07,Tristram_14}. Both components have a dust temperature of roughly 300\,K. The disc-like component was modeled by two Gaussian emitters (a highly 
elongated emitter plus an unresolved source) in the simple model used by \citet{Tristram_14}. This was interpreted as signs for a more complex structure as well 
as for asymmetry in the second component. The orientation and size of the disc-like component coincides with an edge-on warped disc 
seen in maser emission with the help of Very Long Baseline Interferometry traced by H$_2$O masers
\citep[VLBI,][]{Greenhill_03}. It has an outer radius of roughly 0.4\,pc. 
A similar picture emerges for NGC~1068: A hot parsec-scale disc (FWHM $\approx 1.35 \times 0.45\,$pc, $T\approx 800$\,K) was found with 
similar orientation and extent as the H$_2$O maser disc \citep{Greenhill_96,Greenhill_97}.
It is surrounded by warm dust extended in polar 
direction with a FWHM $\approx 3 \times 4\,$pc and a temperature $T\approx$ 300\,K \citep{Jaffe_04,Raban_09}. 
However, in a systematic study of a larger sample of objects a large range of properties of the dust distribution
has been found \citep{Burtscher_13}. 

As a first step to investigate the observational implications and to determine possible morphologies of AGN tori, 
geometrical toy models have been set up 
\citep[e.~g.~][]{Pier_92b,Granato_94,Schartmann_05,Fritz_06,Hoenig_10b,Schartmann_08,Stalevski_12,Siebenmorgen_15}.
Knowledge gained from this effort was used to set up physical models.
These concentrated on the mass supply \citep{Schartmann_09,Schartmann_10} and their dynamical stability \citep{Wada_02,Wada_09}. 
Especially the role of radiation pressure has gained substantial interest in the recent 
years \citep[e.~g.~][]{Krolik_07,Schartmann_11,Dorodnitsyn_12,Wada_12,Schartmann_14,Wada_15,Chan_16,Wada_16} following the 
early work of \citet{Pier_92a}. Alternative theories have postulated dusty winds to completely replace 
AGN tori \citep{Emmering_92,Koenigl_94,Elitzur_06}.
For a more thorough discussion of possible AGN torus models we refer to \citet{Chan_16}.

In this article we will concentrate on the observational appearance of geometrically thin, but warped discs. Although the idea was brought forward 
already more than 25 years ago by \citet{Sanders_89}, it has thus far gotten only very little attention.
Its main advantage is that a significant covering fraction of the sky can be reached without the need of having a large scale height. 
In this context, \citet{Lawrence_10} discuss misaligned discs as possible sources of AGN obscuration.
Maser emission at 22\,GHz has been found in a large number of nuclei in active galaxies \citep[e.~g.~][]{Kartje_99}.
In some of the best-studied cases they appear in warped disc structures, e.~g.~NGC~4258 \citep{Greenhill_95,Herrnstein_96}, NGC~1068 
\citep{Greenhill_96,Greenhill_97}, NGC~3393 \citep{Kondratko_08} and the Circinus galaxy \citep{Greenhill_03}. 
Several physical mechanisms have been suggested to trigger the warping of discs, like e.~g.~radiation pressure induced warping instability
\citep{Pringle_96,Maloney_96}, an orbiting massive object \citep{Papaloizou_98}, stellar torques from a nuclear star cluster \citep{Bregman_12} or 
stellar rings \citep{Nayakshin_05}, misalignment between momentum of infalling material and the spin of the 
central black hole or due to the presence of a binary black hole \citep[e.~g.~][]{Tremaine_14}.

Dust can play an important role in the thermodynamics of the maser pump cycle \citep{Collison_95}. This together with measured inner radii of the maser
emission close to the sublimation radius of dust makes the coexistence of gas and dust
in these systems very likely. 

The main aim of this publication is to do a first investigation of the observational consequences of dusty, warped discs in the (sub-)\,parsec scale 
environment of nearby Seyfert galaxies. 
Our standard model is one realisation of a warped disc 
structure inspired by the properties of the nucleus of the Circinus galaxy and we discuss the effects of varying its main morphological parameters. 
The underlying density distribution of the model and methods used are summarised in Sect.~\ref{sec:setup_methods}.
The typical features of such a warped disc model will be presented in Sect.~\ref{sec:char_appearance}, a comparison with 
the Circinus galaxy is done in Sect.~\ref{sec:circinus_comparison} and we will discuss benefits and shortcomings in Sect.~\ref{sec:discussion}. We summarise our work and draw our final conclusions in Sect.~\ref{sec:conclusions}.

\section{Model setup and methods}
\label{sec:setup_methods}
\subsection{The density distribution}
\label{sec:density_distribution}
\begin{figure}
	\centering	
 	\includegraphics[width=\columnwidth]{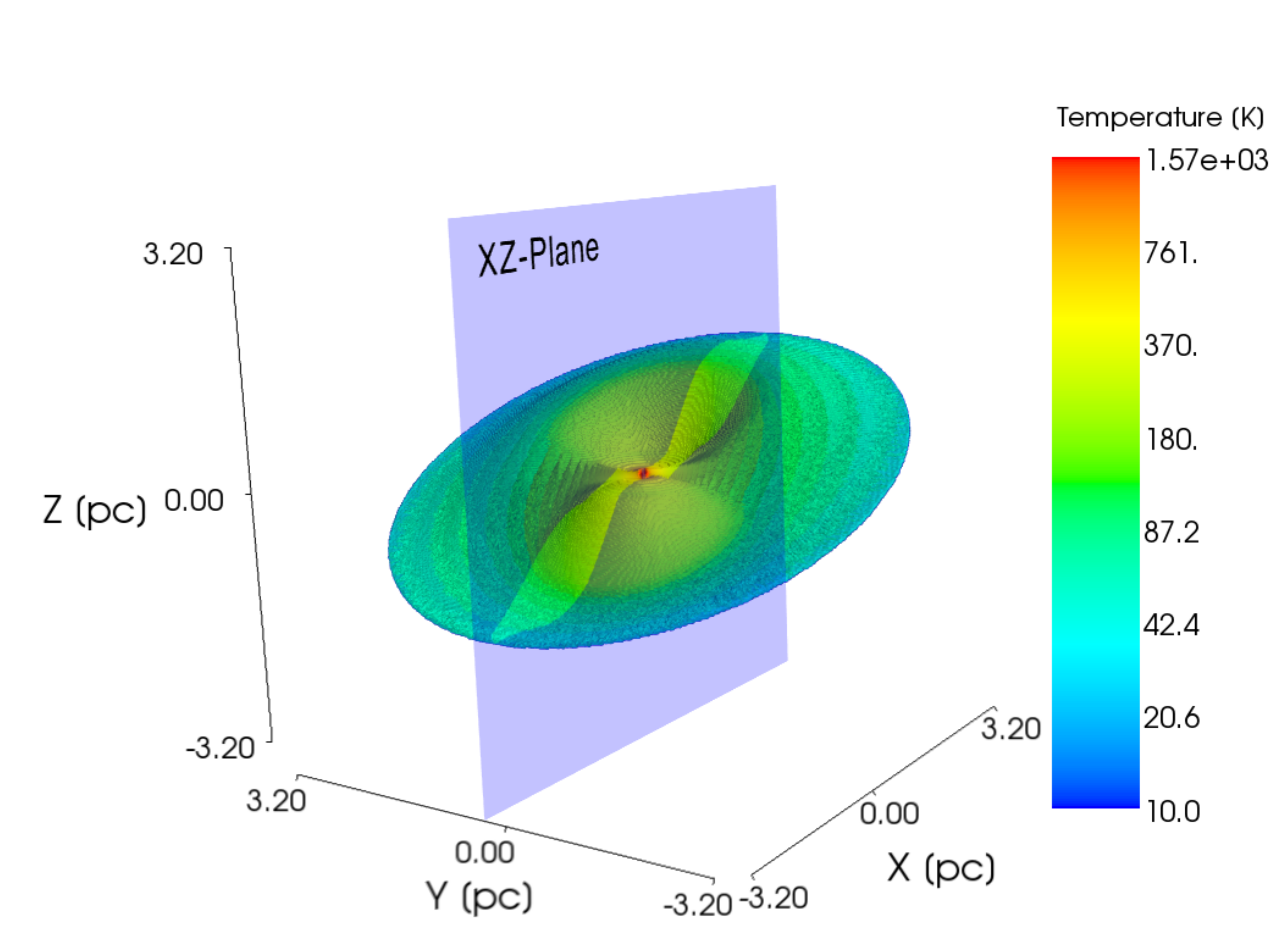}
 	\caption{Semi-transparent 3D temperature distribution of the dust of our standard model. The XZ plane is plotted in transparent blue while the parts where it cuts the dust distribution are coloured in yellow to visualise the warp in x direction. The dust reaches temperature values up to \SI{1560}{\kelvin}. The visible disc is continuously filled with dust.}
 	\label{fig:3Dtemperature_distribution}
\end{figure}
\begin{figure*}
	\centering	
 	\includegraphics[width=\textwidth]{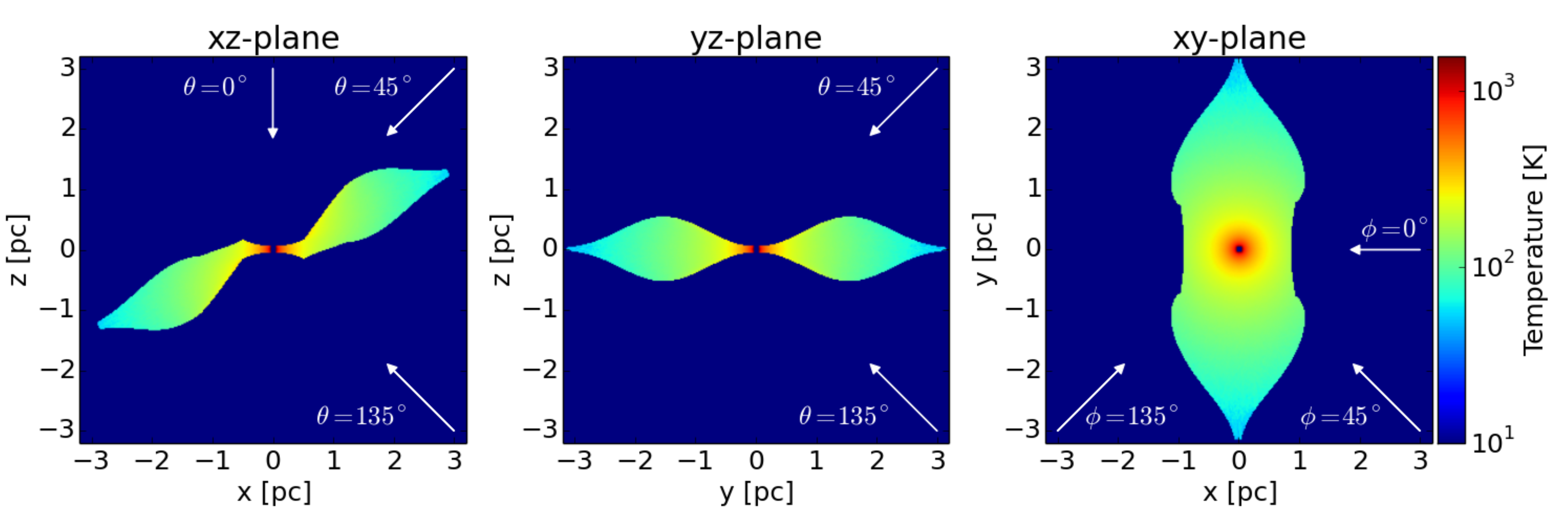}
 	\caption{Cuts through the temperature distribution of the dust of our standard model, taken along the coordinate planes (xz plane: $\theta = 90^\circ$, $\phi = 90^\circ$, yz plane: $\theta = 90^\circ$, $\phi = 0^\circ$, xy plane: $\theta = 0^\circ$, $\phi = 0^\circ$). A logarithmic temperature scale is used where everything with a temperature lower than \SI{10}{\kelvin}, as well as the regions without dust is plotted in dark blue. The dust reaches temperature values up to \SI{1560}{\kelvin}.}
 	\label{fig:temperature_distribution}
\end{figure*}
For the sake of simplicity we adopt a geometrical model with a constant dust density for the disc around the central source. 
The inner radius of the disc is chosen such that the highest temperature to which the dust grains are heated is close to the sublimation temperature, 
which we assume to be \SI{1500}{\kelvin} for our dust mixture (Fig.~\ref{fig:3Dtemperature_distribution}). 
The warping angle is chosen such to resemble the one derived by \citet{Greenhill_03} for the Circinus galaxy.
As maser emission traces only the densest part of the disc, 
the choice of the other parameters of the model are motivated by MIR observations of the 
Circinus Galaxy by~\citet{Tristram_14}, which include the value of the outer radius for which we adopt 3.15\,pc.
Together with the density of the dust, 8 parameters characterise our model. In the following we will explain the geometrical functions we use to build 
up the dust distribution. We first describe what we call the {\it flat model} and then explain how we apply the warp by changing the coordinate grid used to calculate the flat model.

The cross section of the flat model is a wedge which has been modified to have a concave shape at small radii and which turns over at large radii (the cross section of the flat model is identical to the cut shown in the middle 
panel of Fig.~\ref{fig:temperature_distribution}).  
We adopt a concave shape close to the source to resemble hydrostatic equilibrium configurations but also to have directly illuminated dust closer to the source. To generate the flat model we first calculate the height of the disc according to the following function
\begin{equation}
	d_\mathrm{h}(r_\mathrm{cyl}) = d_\mathrm{h}(0) + \frac{4}{5} \, r_\mathrm{cyl} \, \tan (\xi) \, ,
\end{equation} 
where $d_\mathrm{h}$ is half of the disc thickness in $z$-direction as a function of the cylindrical radius $r_\mathrm{cyl}$, i.e., the distance to the $z$-axis, and $\xi$ is the opening angle of the wedge. Furthermore, we implement a minimal disc height of $d_\mathrm{h}(0)$ in the inner part to account for limitations in resolution. To get a concave shape of the disc close to the source, we make the opening angle $\xi$ a function of the radius. The following function decreases the opening angle for values of $r_\mathrm{cyl}$ close to the inner ($r_\mathrm{in}$) and outer ($r_\mathrm{out}$) boundary of the disc
\begin{equation}
	\xi (r_\mathrm{cyl}) = \sin \left( \frac{r_\mathrm{cyl}-r_\mathrm{in}}{r_\mathrm{out}/2-r_\mathrm{in}} \right)\, .
\end{equation}
To flatten the disc and get closer to the shape expected from hydrodystatic equilibrium we finally apply the following function
\begin{equation}
	d_\mathrm{h}^{\mathrm{final}}(r_\mathrm{cyl}) = d_\mathrm{h}(r_\mathrm{cyl}) \, \sin \left( \frac{r_\mathrm{cyl}-r_\mathrm{in}}{r_\mathrm{out}-r_\mathrm{in}}+\frac{\pi}{2} \right)^8 \, .
\end{equation}

To warp the disc we change the $z$-coordinates of our coordinate grid, which we use to calculate the flat model described above.
We warp the disc in $z$-direction with maximum amplitude in the $xz$-plane from an inner radius which we denote by $r_\mathrm{win}$ to an outer 
radius $r_\mathrm{wout}$. 
The new coordinate $\tilde{z}$ is given as follows
%
\begin{equation}
	\tilde{z}(x,y,z) = z + z_\mathrm{w} \sin \left( \frac{\pi}{2} \, \zeta \, \frac{x-\sign (x) \, x_\mathrm{di}}{\left( r_\mathrm{wout}-r_\mathrm{win}\right)} \right) \, ,
\end{equation}

where $z_\mathrm{w}$ parametrises the largest possible dislocation.
We smoothly decrease the warp as we approach the $y$-axis to end up with no warp at all in that direction, which we parametrise by  

\begin{equation}
	x_\mathrm{di} (x, y) = \frac{r_\mathrm{win}}{r_\mathrm{cyl}}\,x\, .
\end{equation} 
Furthermore, $\sign (x)$ denotes the sign of the coordinate $x$.
Finally, we cut out the dust lying outside the spherical shell limited by the radii $r_\mathrm{in}$ and $r_\mathrm{out}$ and continue the non-warped disc smoothly 
outside the warp-region defined by $r_\mathrm{win}$ and $r_\mathrm{wout}$. 
The parameters we use for our calculations are given in table~\ref{tab:parameters}. Furthermore, a visualisation of the 3D temperature distribution
for our standard model is shown in Fig.~\ref{fig:3Dtemperature_distribution}. 
Its outer contour corresponds to the shape of the dust distribution.

The restriction to a model with constant density allows us to circumvent limitations in the achievable spatial resolution and optical depth of the modelling. 
For a disc in hydrostatic equilibrium around a central supermassive black hole a steep density gradient around the equatorial plane would be expected \citep[e.~g.~][]{Pringle_81}.
This means that the total mass given in Table~\ref{tab:parameters} is probably underestimating the real mass.
However, the inner part of our distribution's morphology roughly follows the outer shape of such a flared disc geometry.   
\begin{table}
	\centering
 	\caption{Overview over the parameters used for the simulation of our models. 
 	    The middle column corresponds to our standard model and 
 	    the right column depicts those parameters varied in a parameter study, resulting
 	    in a total of 16 models.
 		$M_\mathrm{BH}$ is the mass of the central supermassive black hole,
 		$L_\mathrm{disc}$ the luminosity of the central heating source,
 		$M_\mathrm{dust}$ the total amount of dust,
 		$\rho_\mathrm{dust}$ the dust density,
 		$r_\mathrm{in}$ the inner radius of the dust distribution,
 		$r_\mathrm{out}$ the outer radius,
 		$d_\mathrm{h}(0)$ half of the minimal disc height,
 		$r_\mathrm{win}$ the inner radius of the warp,
 		$r_\mathrm{wout}$ the outer radius of the warp,
 		$z_\mathrm{w}$ and $\zeta$ additional parameters defining the warp,
 		$n_\mathrm{phot}$ number of photon packages for the Monte Carlo (MC) run,
 		$n_\mathrm{phot\_scat}$ number of photon packages for the scattering MC runs and
 		$v_\mathrm{c}$ the cell volume used.
 	}
 	\label{tab:parameters}
 	\begin{tabular}{lcc} 
 		\hline
 		\hline	
 		Parameter & Standard model & Parameter variation \rule{0pt}{2.6ex}\rule[-0.9ex]{0pt}{0pt}\\ 		
 		\hline
 		$M_\mathrm{BH}$ & $1.7 \times 10^6 \, \mathrm{M_\odot}$ &  \rule{0pt}{2.6ex} \\
 		$L_\mathrm{disc}$ & $1.64 \times 10^{10} \, \mathrm{L_\odot}$ &  \\
 		$M_\mathrm{dust}$ & $170 \, \mathrm{M_\odot}$ & \\
 		$\rho_\mathrm{dust}$ & $6.8 \times 10^{-22} \, \text{g~cm}^{-3}$ & \\
 		$r_\mathrm{in}$ & $0.05 \, \mathrm{pc}$ & \\
 		$r_\mathrm{out}$ & $3.15 \, \mathrm{pc}$ & \\
 		$d_\mathrm{h}(0)$ & $0.05 \, \mathrm{pc}$ & \\
 		$r_\mathrm{win}$ & $0.5 \, \mathrm{pc}$ & $0.25 \, \mathrm{pc}$ \\
 		$r_\mathrm{wout}$ & $1.2 \, \mathrm{pc}$ &  $2.4 \, \mathrm{pc}$\\
 		$z_\mathrm{w}$ & $1.0 \, \mathrm{pc}$ & $0.5 \, \mathrm{pc}$\\
 		$\zeta$ & $0.4$ & $0.8$\\
 		\hline
 		$n_\mathrm{phot}$ & $10^8$ & \rule{0pt}{2.6ex} \\
 		$n_\mathrm{phot\_scat}$ & $10^5$ & \\
 		$v_\mathrm{c}$ & $8 \times 10^{-6} \, \mathrm{pc}^3$ &  \\
 		\hline
 	\end{tabular}
\end{table}

\subsection{The radiative transfer approach}
\label{sec:rtapproach}

For our simulations we use the 3D dust continuum radiative transfer capabilities of {\sc RADMC-3D}\footnote{\url{http://www.ita.uni-heidelberg.de/~dullemond/software/radmc-3d/}} (version 0.39\_17\_03\_15). The code is based on the Monte Carlo approach and allows us to first compute the dust temperature. We can then use this temperature distribution as an input for the calculation of the SEDs, after specifying a central source and the characteristics of the dust grains. {\sc RADMC-3D} computes the dust temperature by dividing the total luminosity of the source into $n_\mathrm{phot}$ packages, which are then emitted by the source one at a time. Moving through the coordinate grid they may scatter or be absorbed by the dust grains according to the absorption and emission coefficients specified. In going through the grid each photon package entering a cell increases the cell's energy and in turn also the dust temperature associated to that cell. 
The final dust temperature is obtained once all photon packages have escaped and an equilibrium has been reached. As an approximation we only use the isotropic scattering mode of the code. Furthermore, the only effect scattering can have is changing the direction of a photon. In a second step we can then use the ray-tracing capabilities of {\sc RADMC-3D} to make images and spectra.
For these computations, a transfer equation containing a scattering source function has to be integrated for each pixel, where the integration goes along the one-dimensional ray belonging to that pixel. However, to do that one has to know the scattering source function which {\sc RADMC-3D} computes through a so-called {\it scattering Monte Carlo} run for a single-wavelength. The number of photon packages we use for the scattering Monte Carlo simulation is $n_\mathrm{phot\_scat} = 10^5$ for both the creation of images and spectra,  if not explicitly stated otherwise. 

For our simulations of the warped model we use a three-dimensional Cartesian coordinate grid consisting of 320 cells in each direction ranging from $-3.2$ to $3.2 \, \mathrm{pc}$. We use an equally spaced grid which results in a cell volume of $v_\mathrm{c}=8 \times 10^{-6} \, \mathrm{pc}^3$. We also performed simulations for the flat model, which we could implement in a two-dimensional spherical coordinate system exploiting the symmetry. 
This enabled us to do simulations at a higher spatial resolution. A comparison with our standard Cartesian grid simulations showed that the results are sufficiently converged for our conclusions. The results for the flat model shown in Sect.~\ref{sec:flat_case} are based on a uniform grid with 320 values in radial direction ($R \in [0.01,3.2]$pc) and 192 values for the polar angle ($\theta \in [0, \pi ]$). The flat model is generated with the same parameters and procedure as we use in the warped case just without applying the warp. For the computation of the SEDs we choose a logarithmic wavelength grid with values of $\lambda \in [0.001,2000]\,\text{\SI{}{\micro\meter}}$. The range has been divided into three regions, 
with 20 logarithmically-spaced values in the interval $[0.001,2[\,\text{\SI{}{\micro\meter}}$, 140 in $[2,25[\,\text{\SI{}{\micro\meter}}$ and 30 in $[25,2000]\,\text{\SI{}{\micro\meter}}$. An overview over the input parameters we use is given in table~\ref{tab:parameters}.     
\subsection{Central source and dust characteristics}
\begin{figure}
	\centering	
 	\includegraphics[width=0.95\linewidth]{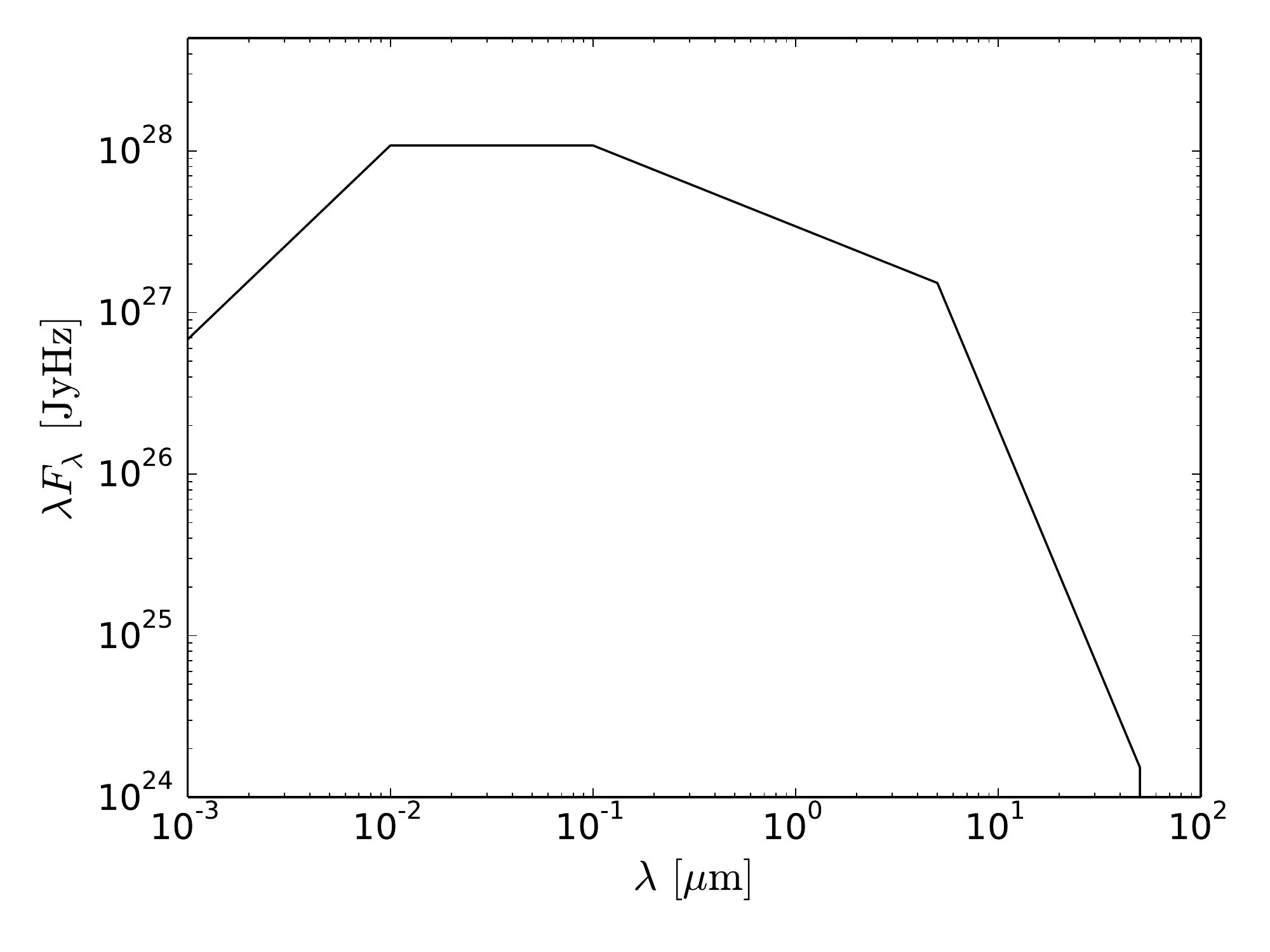}
 	\caption{SED of the central heating source, which is assumed to radiate according to a broken power law. Normalisation is for a distance of 1~pc.}
 	\label{fig:Source_SED}
\end{figure}
We assume the dust to be heated by a central accretion flow surrounding a black hole and approximate its emission by an isotropic, 
central, point-like energy source. 
We used isotropic emission in order to prevent us from having to make any specific assumption about the orientation of the accretion disc. 
As it turns out, our general results are not affected by this choice (see discussion in Sect.~\ref{sec:discussion}).
We set the central source luminosity to $1.64 \times 10^{10}\,\mathrm{L_\odot}$ which corresponds 
to 30\% of the Eddington luminosity for a black hole mass of $1.7 \times 10^6\,\mathrm{M_\odot}$ as derived for the Circinus galaxy by maser disc kinematics \citep{Greenhill_03}. This is in accordance with the findings of \citet{Tristram_07} who derive a luminosity of $10^{10}\,\mathrm{L_\odot}$.
The SED of the source is shown in Fig.~\ref{fig:Source_SED} and is assumed to be composed of different power laws with spectral indices chosen as in~\citet{Stalevski_12}, i.e.  
\begin{equation}
	\lambda F_\lambda \propto \begin{cases}
									\lambda^{1.2} \qquad	&0.001  \leq \lambda \, \text{\si\micro\si\meter}^{-1}\, \leq 0.01\\
									\lambda^{0} 		&0.01  < \lambda \, \text{\si\micro\si\meter}^{-1}\, \leq 0.1\\
									\lambda^{-0.5}		&0.1  < \lambda \, \text{\si\micro\si\meter}^{-1}\, \leq 5\\
									\lambda^{-3}		&5  < \lambda \, \text{\si\micro\si\meter}^{-1}\, \leq 50 \,.
								\end{cases}
\end{equation}
A typical galactic dust model is used, similar to the one presented in \citet{Schartmann_05}. We set up a mixture of 62.5\% silicate grains and
47.5\% of graphite grains (taking the two possible orientations of the 
optical axis into account),
distributed according to a grain size distribution following
\citet{Mathis_77}. 
Optical constants are taken from \citet{Draine_84}, \citet{Laor_93} and \citet{Weingartner_01}. 
With these ingredients, we calculate an average dust model
following the approach of \citet{Wolf_03b}.
\section{Characteristic appearance}
\label{sec:char_appearance}
In the following we present the main results of our work obtained using the radiative transfer code {\sc RADMC-3D} with the density distribution as described 
in section~\ref{sec:density_distribution}. We will interpret images for different lines of sight and wavelengths as well as the corresponding SEDs to characterise 
our standard model and discuss consequences of parameter variations. All SEDs we show in this paper include both the dust re-emission spectra as well as the radiation from the central source, which is added by {\sc RADMC-3D} in a post-processing step (i.e.\,, to the image first rendered without the source, the flux obtained by a ray tracing from the source to the observer is added at the correct position, only taking extinction into account). 

\subsection{The temperature distribution}
The temperature distribution of our standard model is shown in Fig.~\ref{fig:3Dtemperature_distribution} and Fig.~\ref{fig:temperature_distribution}, 
where the temperature of all cells which contain no dust has been set to zero. The highest temperature to which the dust is heated is \SI{1560}{\kelvin} 
which is approximately the assumed sublimation temperature of the dust. 
The three panels in Fig.~\ref{fig:temperature_distribution} show cuts through the temperature distribution along the coordinate planes. In the left panel, which shows a cut along $y=0$, one can see the effect of the warp 
in $z$-direction depending on the $x$-coordinate as explained in section~\ref{sec:density_distribution}. The flat disc inside the start of the warp at $r_\mathrm{win}$, as well as the 
final angle w.r.t the $x$-axis can clearly be seen. Due to the warping of the disc, the temperature distribution is clearly asymmetric as only one 
side can be illuminated directly from the central source.

Rotating the cut towards the $yz$-plane, the warp is slowly reduced to zero. The cut along the $yz$-plane is plotted in the middle 
panel and shows no difference to a cut one would obtain for the flat model. Finally, in the right panel the temperature distribution of the $xy$-plane is shown. In this plot one can also 
observe the effect of the warp. Without a warp the plot would show a spherical shell limited by the inner and outer radii of the dust distribution. However, here we see that the dust 
distribution is warped out of the plane towards negative -- respectively positive $z$ values depending on the $x$-coordinate. The effect of the flat inner disc can also be seen as most 
of the dust heated to the highest temperatures still lies inside this plane. 

\subsection{Images at various lines of sight}
In the following we show images of our standard model for different inclination ($\theta$) and rotation angles ($\phi$) taken at a wavelength of \SI{12}{\micro\meter}. The angles  specify the location of the observer, where the inclination is measured from the $z$-axis and may be characterised by two vectors. One of these vectors points along the negative $z$-axis, corresponding to $\theta = 0^\circ$, the other one, associated with $\theta = 90^\circ$, lies in the $z=0$ plane. The rotation angle rotates the location of the observer clockwise around the $z$-axis, i.e., the model is rotated counter-clockwise w.r.t the observer. Therefore, for a flat model one would obtain a face-on view for an angle of $\theta = 0^\circ$, corresponding to an extreme Seyfert~1 case. An edge-on view and thus an extreme Seyfert~2 case would be seen for an angle of $\theta = 90^\circ$. In our warped model the geometry is somewhat more elaborate and it is thus not possible to define an explicit edge-- or also face-on views as there are always parts of the dust which may be directly illuminated or obscure the central source. 

\begin{figure*}
	\centering
  	\includegraphics[width=\textwidth]{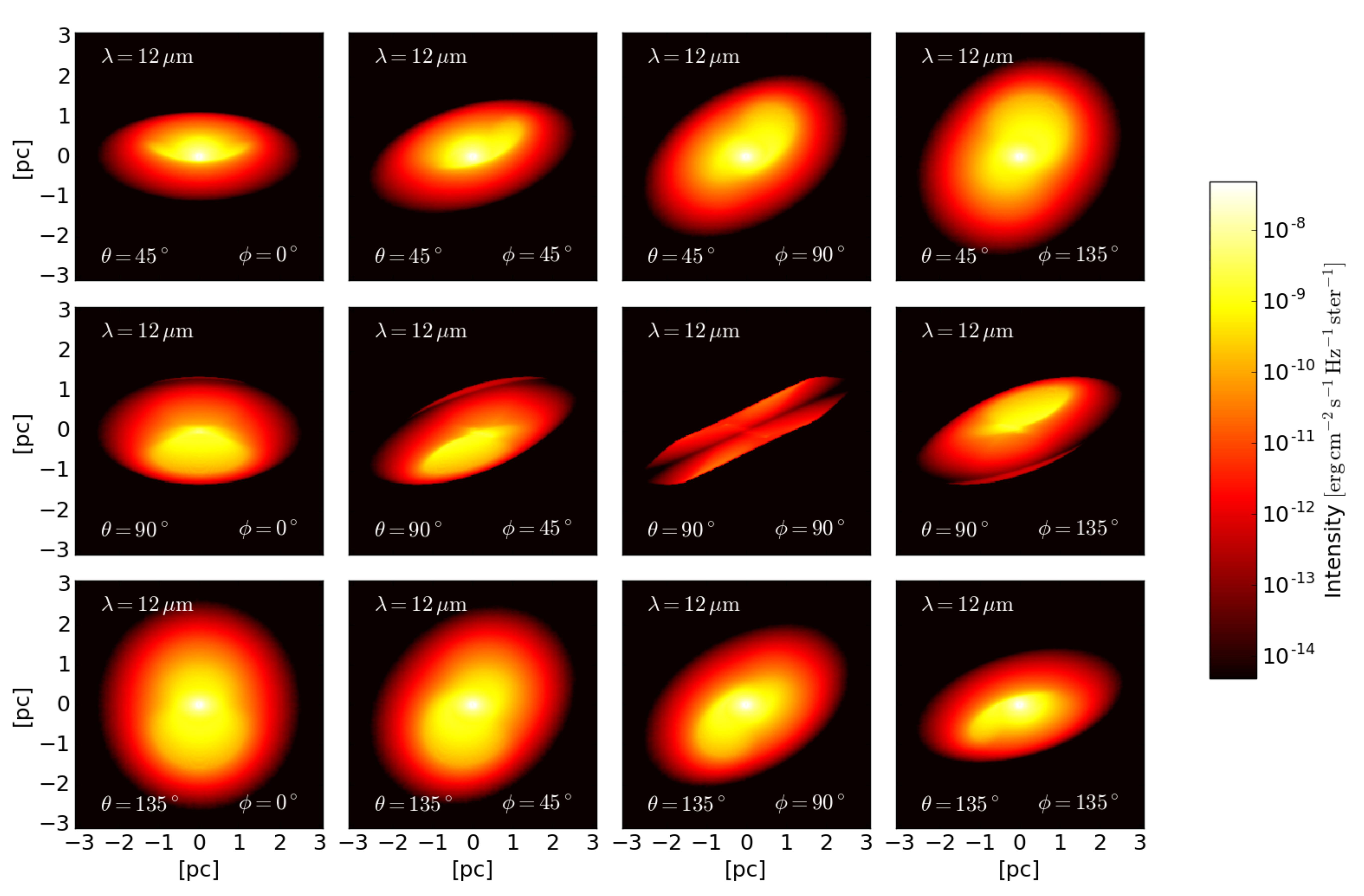}
  	\caption{Images of the intensity for different lines of sight towards our standard model. The images are taken at a wavelength of $\lambda =$\,\SI{12}{\micro\meter}. The inclination w.r.t. the $z$-axis is denoted by $\theta$ and the rotation angle around it by $\phi$. $\theta\, , \phi = 0^\circ$ is a line of sight along the negative $z$-axis and $\theta = 90^\circ$, $\phi = 0^\circ$ one along the negative $x$-axis, i.e., the coordinate system is such that it fits the one chosen for the temperature distribution (Fig.~\ref{fig:temperature_distribution}). All parts with a lower intensity than the minimum shown in the legend are plotted in black.}
  	\label{fig:Images_incl}
\end{figure*}

In Fig.~\ref{fig:Images_incl} we show images at a wavelength of \SI{12}{\micro\meter} for 
inclinations of $\theta = 45,\, 90,\, \text{ and } 135^\circ$ and rotation angles of $\phi = 0,\, 45,\, 90 \text{ and } 135^\circ$. 
The models possess point symmetry, which means that we get the same image for angles of $\theta = 45^\circ$, $\phi = 0^\circ$ and 
$\theta = 135^\circ$, $\phi = 180^\circ$ up to a reflection at the horizontal axis.
The upper row shows an inclination angle (slightly) outside the opening angle of the disc. Hence, a large fraction of the directly illuminated part of the disc 
is visible for all rotation angles $\phi$. The left-most image is viewed along the direction of the strongest warp with an almost grazing line of 
sight w.~r.~t.~the warp in positive $z$-direction (compare to Fig.~\ref{fig:3Dtemperature_distribution} and \ref{fig:temperature_distribution}). 
With increasing rotation angle $\phi$, 
a larger and larger fraction of the directly illuminated surface of this upper warp is visible and a characteristic pear-shaped structure appears. 
The same behaviour is observed when the model is viewed from 
below the $z=0$ plane (lower row). Due to the point symmetry, the lower row of 
Fig.~\ref{fig:Images_incl} can be thought of a continuation of the rotation shown in the upper row.
All images appear elongated in direction of the main warp. 

\begin{figure}
	\centering	
 	\includegraphics[width=200px]{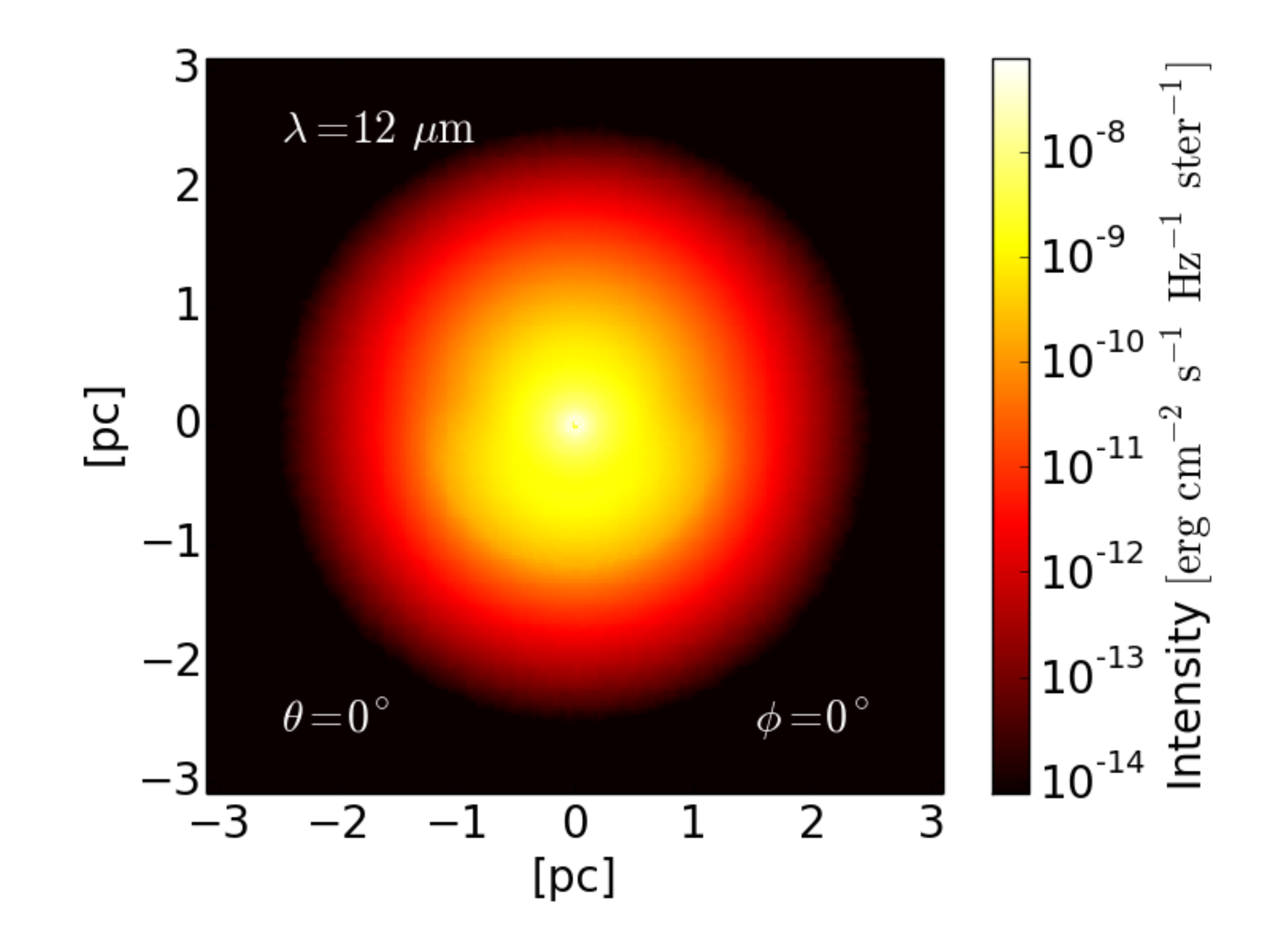}
 	\caption{Face on view of the intensity of our standard model for an observer characterised by $\theta = 0^\circ$, $\phi = 0^\circ$ at a wavelength of \SI{12}{\micro\meter}.}
 	\label{fig:image_face_on}
\end{figure}

Significant differences are visible for lines of sight within the opening angle of the 
disc. This is demonstrated for an inclination angle of $90^\circ$ in the middle row of 
Fig.~\ref{fig:Images_incl}. In this case, the central source is invisible for all
rotation angles. The morphology of the MIR images changes significantly: from a characteristic cone-like appearance when viewed in direction of the warp (left panel)
to a more and more {\it disc-like} appearance and to the asymmetric {\it pear}-shaped 
structure. 
The latter arises from the different curvatures of the warping in positive and negative 
$z$-direction and is even visible in the face-on view (Fig.~\ref{fig:image_face_on} at an inclination angle of $\theta = 0^\circ$).


\subsection{Spectral energy distributions}

\begin{figure*}
	\centering
  	\subfloat{\includegraphics[width=0.3\textwidth]{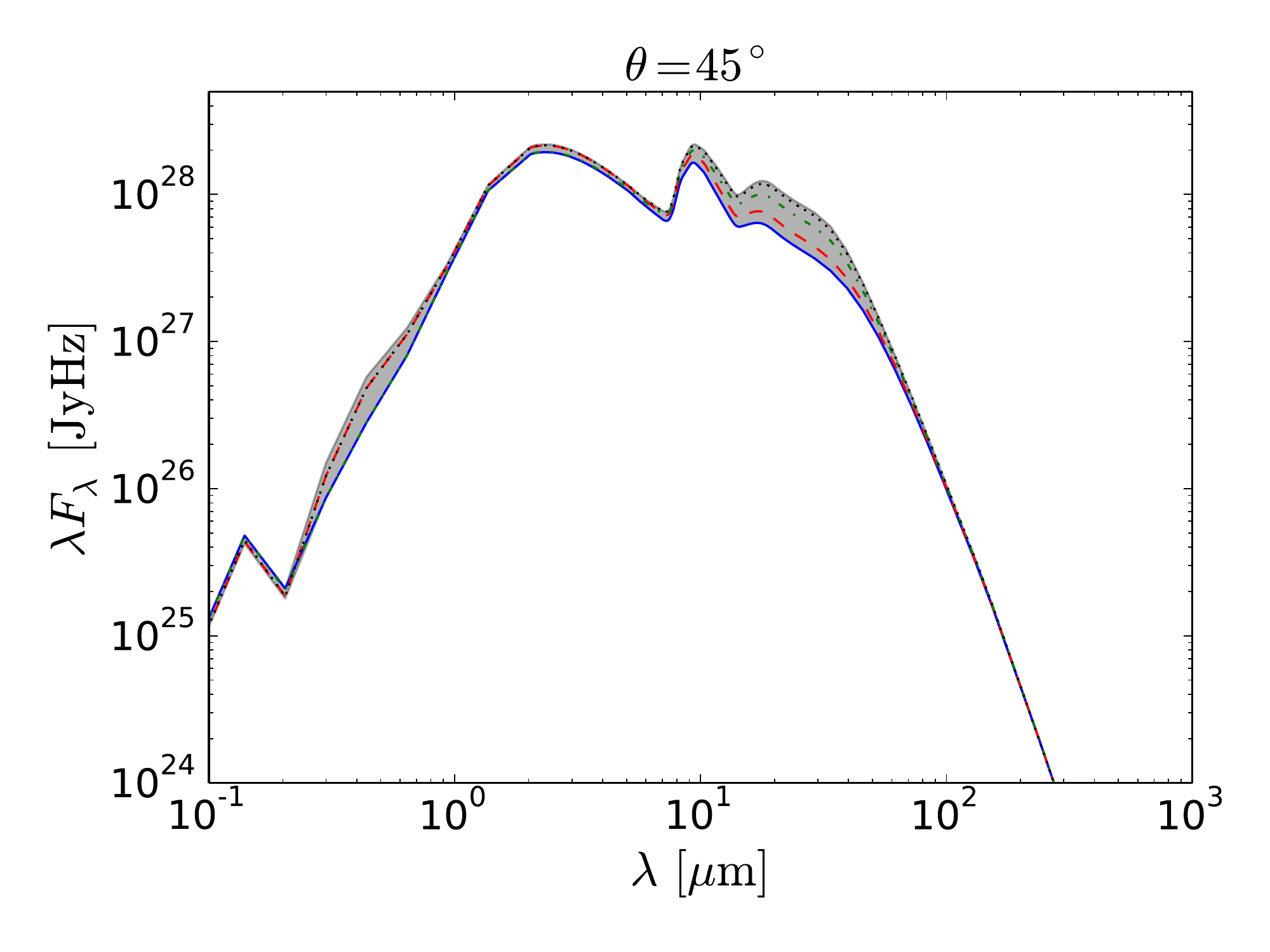}\label{fig:SED_incl_45}}
	\hspace{4pt}
  	\subfloat{\includegraphics[width=0.3\textwidth]{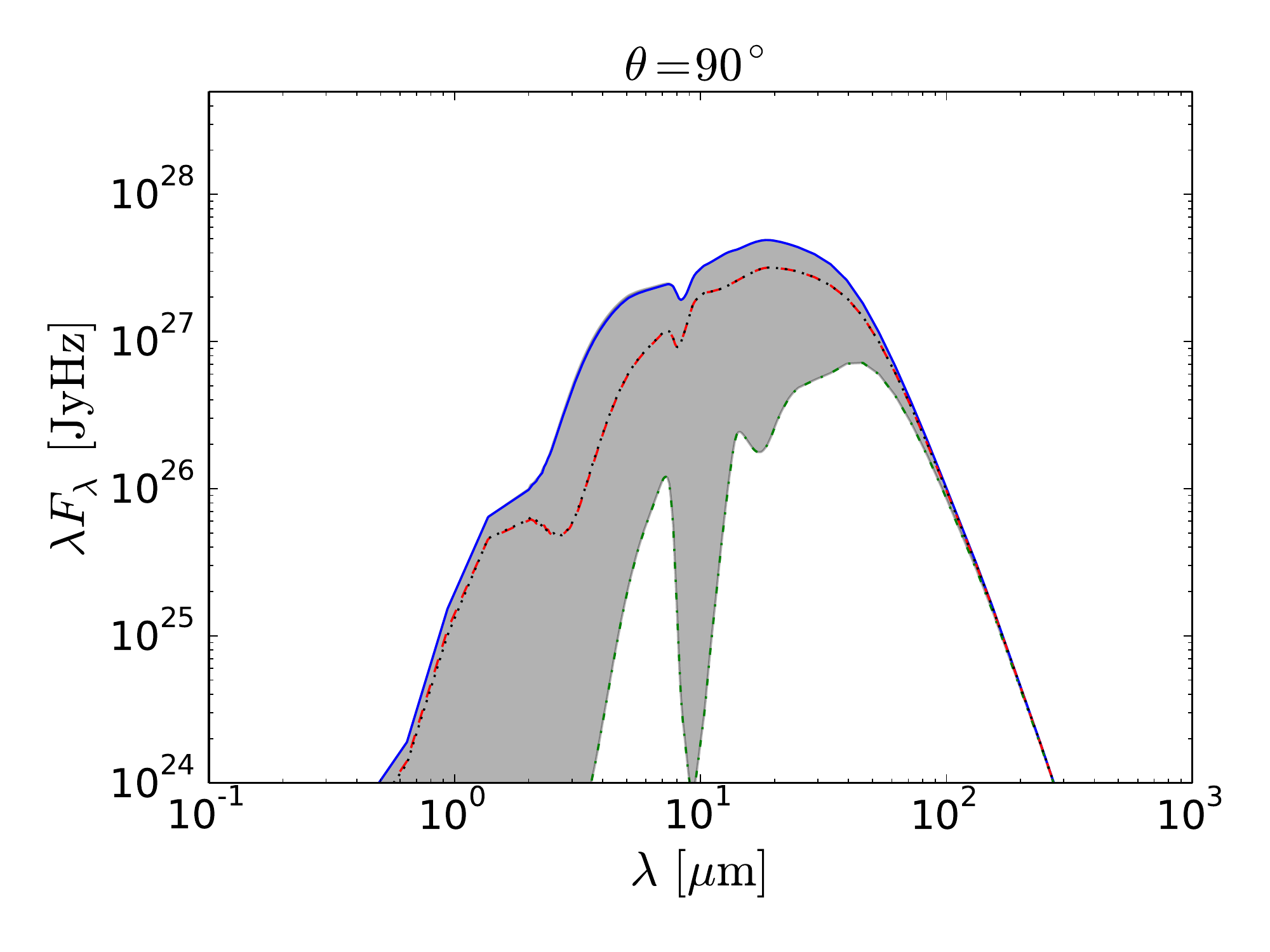}\label{fig:SED_incl_90}}
  	\hspace{4pt}
  	\subfloat{\includegraphics[width=0.3\textwidth]{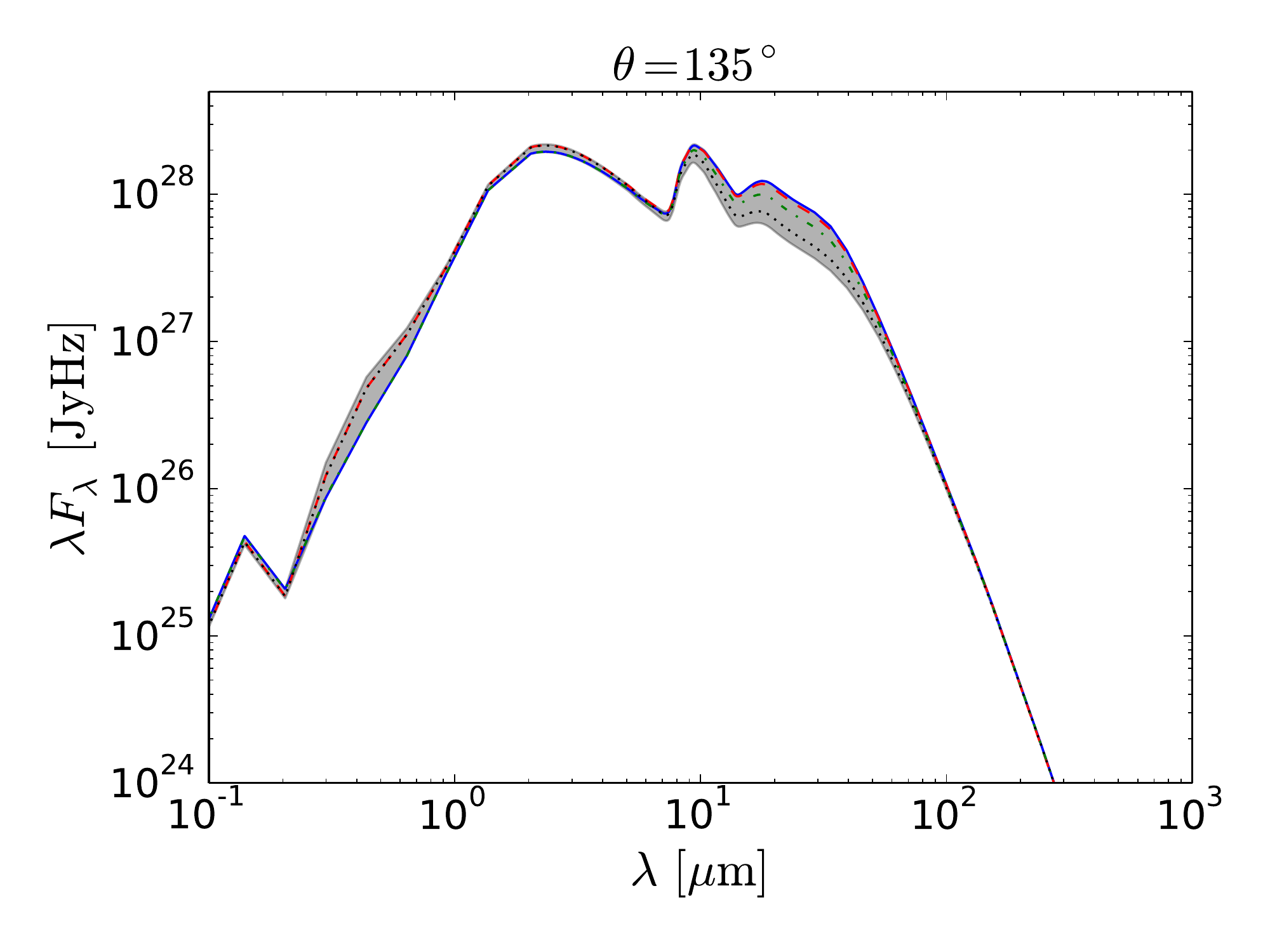}\label{fig:SED_incl_135}}
  	\caption{Dependence of the SEDs of our standard model on different lines of sight for three inclination angles $\theta = 45, 90 \text{ and } 135^\circ$. Shaded is the region between the maximal and minimal values obtained for different values of $\phi$ lying in the range $[0^\circ,180^\circ]$ with a step size of 10 degrees. Some specific results are also plotted, where the solid blue line corresponds to $\phi = 0^\circ$, dashed red to $45^\circ$, dot-dashed green to $90^\circ$ and dotted black to $135^\circ$. The SEDs are normalised to a distance of $1\,\text{pc}$.}
  	\label{fig:SED_incl}
\end{figure*}

In Fig.~\ref{fig:SED_incl} we show the SEDs of our standard model 
corresponding to the position of the observers as chosen for the images in Fig.~\ref{fig:Images_incl}. All SEDs shown in this section are normalised to a distance of $1\, \mathrm{pc}$. 

As already mentioned during the discussion of the images (Sect.~\ref{fig:Images_incl}), an inclination angle of $\theta=45^\circ$ allows a 
direct view of the central directly 
illuminated part for all rotation angles as well as the cold dust in the outer part of the 
warped disc, resulting in a comparatively broad IR SED. In addition, the characteristic silicate features at \SI{9.7} 
and \SI{18.5}{\micro\meter} are visible in emission only. As already discussed in \citet{Granato_94}, all 
viewing angles outside the dusty disc or torus will result in very similar 
SEDs. This is confirmed by the various lines in the last panel of Fig.~\ref{fig:SED_incl}. 
Due to the point symmetry of the model, the shaded region which summarises the range of phi angles is identical for the two 
$\theta$ angles.
Much larger variations are visible for the $\theta=90^\circ$ case. The deepest absorption feature 
is seen for lines of sight perpendicular to the warp direction (green dot-dashed line, $\phi=90^\circ$), 
due to the large column density along the line of sight and thus a high extinction rate.
The feature depth decreases when rotating the model towards a view along the direction
of the strongest warp (blue solid line, $\phi=0^\circ$) as then the largest fraction of the directly illuminated inner surface
of the disc is exposed. The latter leads to a silicate feature in {\it self-absorption} 
\citep{Henning_83}. This means it is combined with an emission feature from these exposed 
surfaces, leading to a narrower silicate feature in absorption, which is slightly 
shifted to shorter wavelengths (see discussion in Sect.~\ref{sec:circinus_comparison}).

\subsection{Parameter dependencies of the model}
\label{sec:par_study}

\begin{figure*}
	\centering
  	\subfloat{\includegraphics[width=0.3\textwidth]{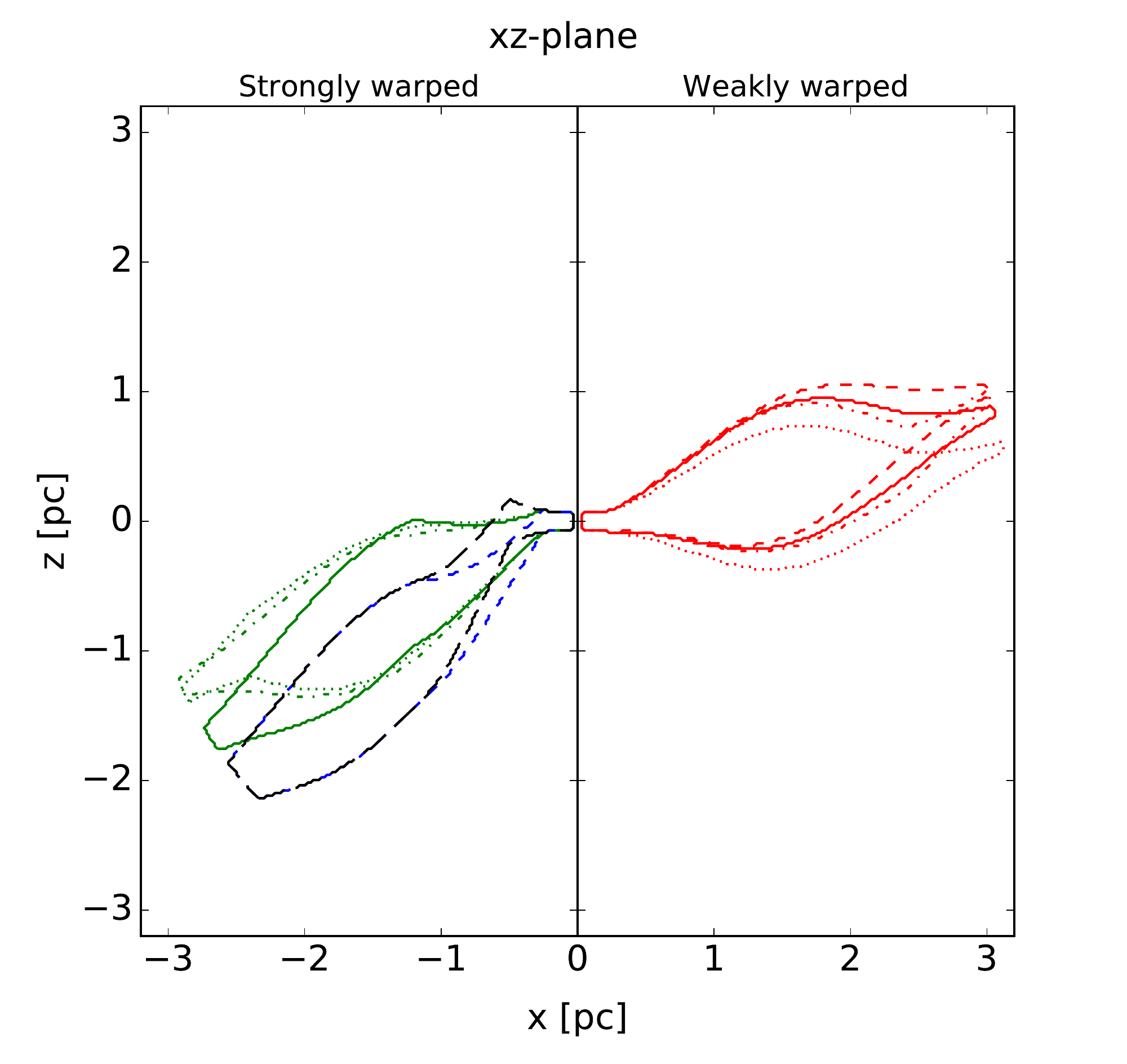}\label{fig:contour_lines_xz_plane}}
	\hspace{4pt}
  	\subfloat{\includegraphics[width=0.3\textwidth]{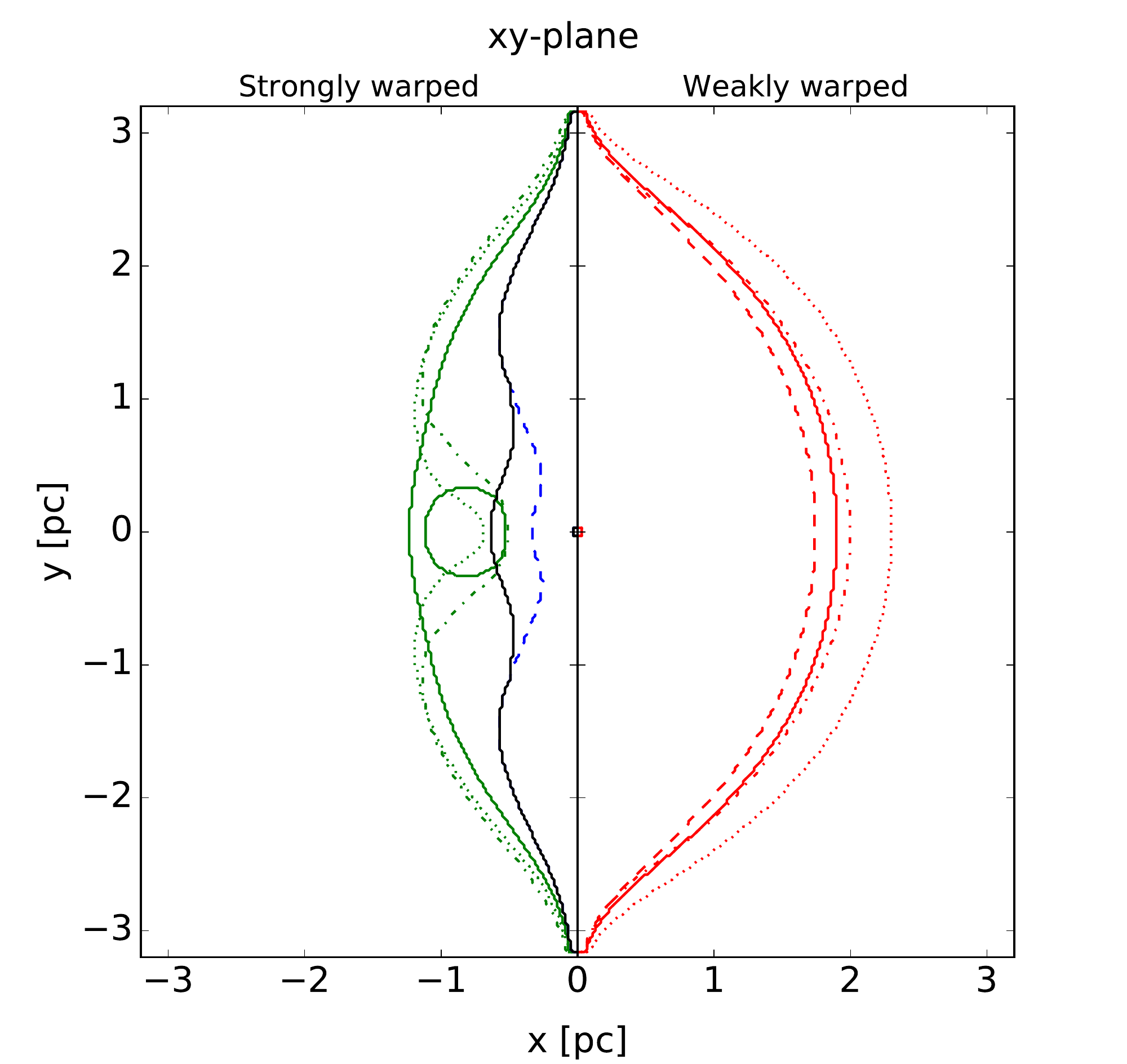}\label{fig:contour_lines_xy_plane}}
	\hspace{4pt}
  	\subfloat{\includegraphics[width=0.3\textwidth]{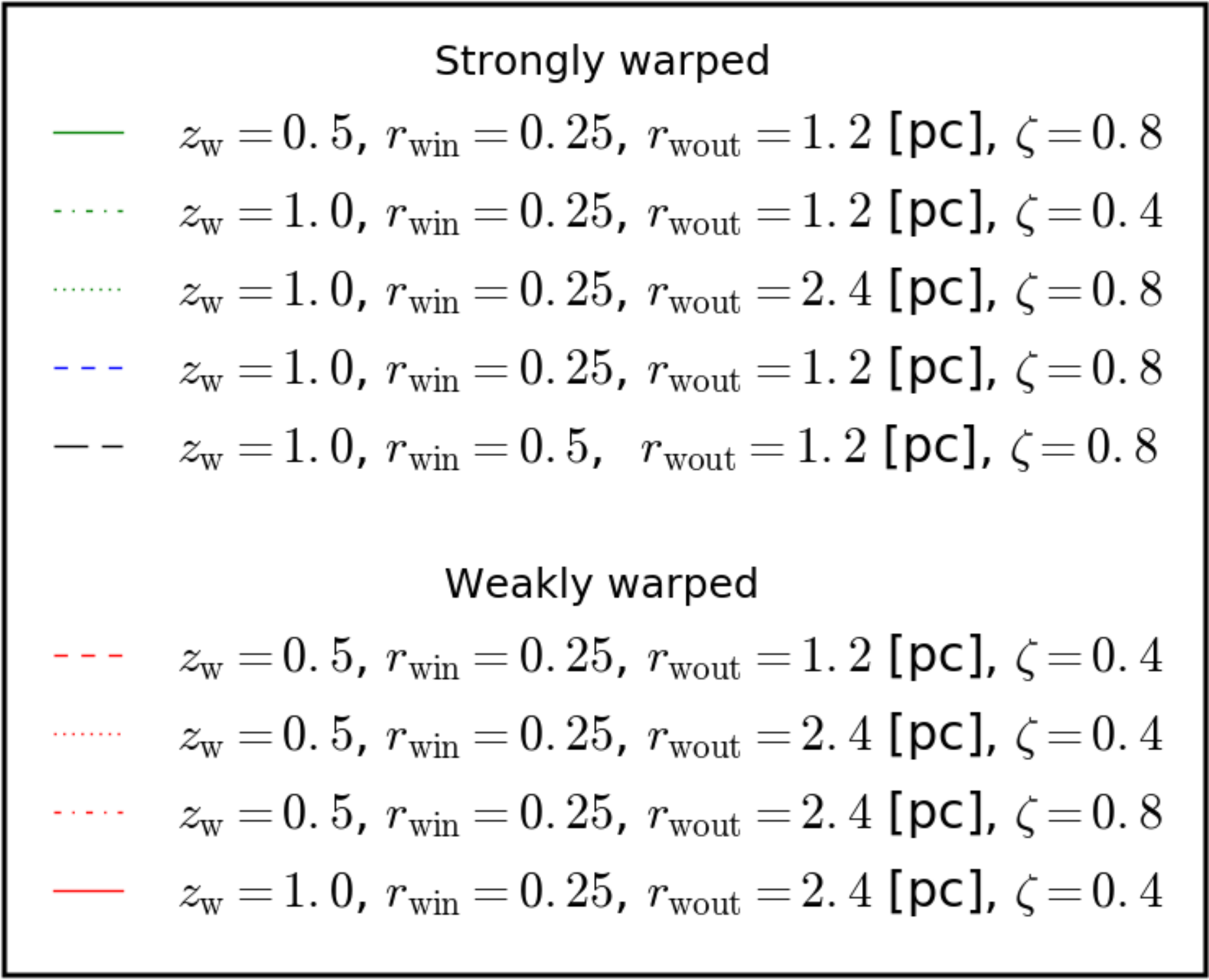}\label{fig:legend}}

  	\caption{Limiting density contours within the $xz$ plane (left panels) and $xy$ plane (right panels) for part of the models of our
  	parameter study (line styles as given in the legend).
  	In the left half of the panels the models 
  	which are classified as strongly warped are plotted whereas the right half of the panels show the weakly warped models. Changing $r_\mathrm{win}$ has 
  	only minor effect on the morphology and observables. Hence only one example is shown (see text for details).
  	}
  	\label{fig:contour_lines}
\end{figure*}
\begin{figure*}
	\centering
	\includegraphics[width=0.7\textwidth]{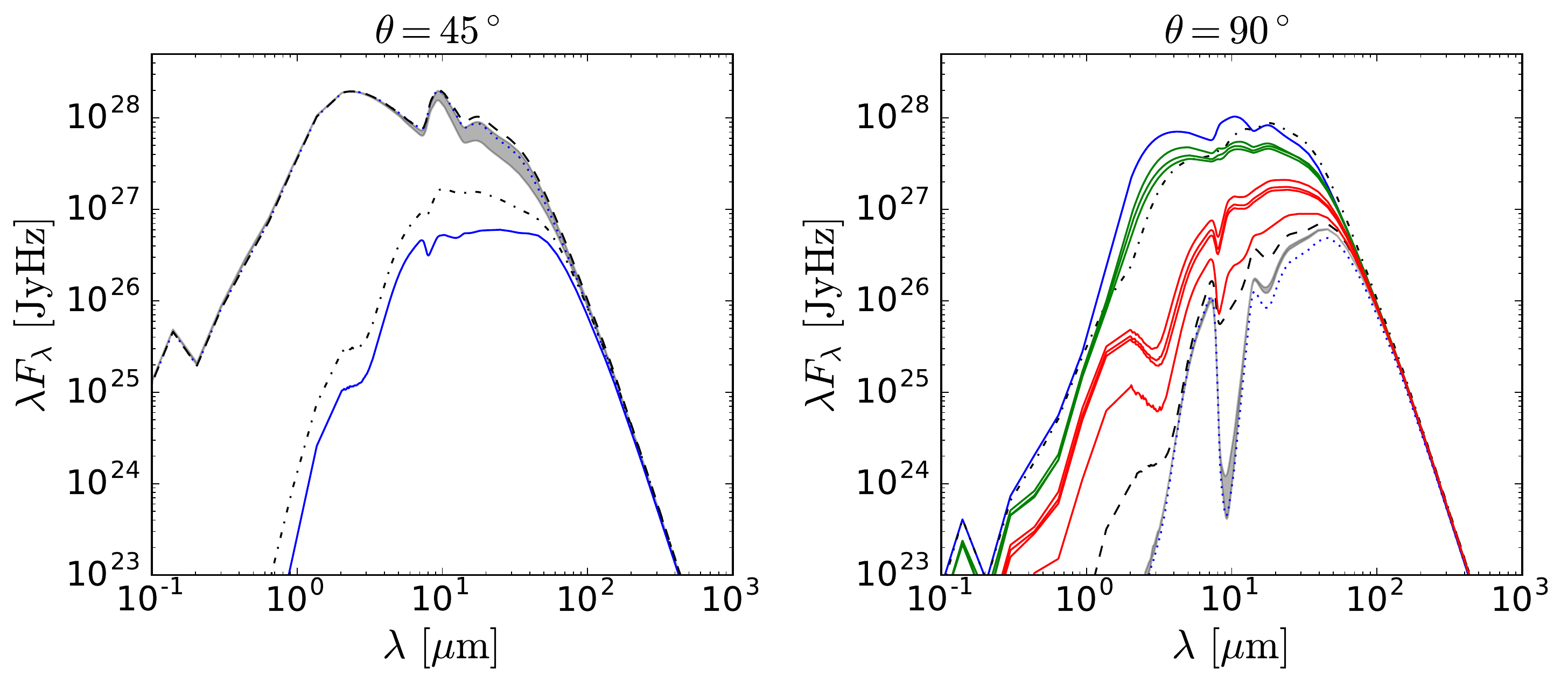}
  	\caption{SEDs of the different models of our parameter study for two inclinations, $\theta = 45^\circ$ and $\theta = 90^\circ$. Shaded in grey are the 
  	SEDs of all weakly and strongly warped models for rotation angles of $\phi = 0^\circ$ and $\phi = 90^\circ$ ($\phi = 90^\circ$ only in the right panel) except the two models characterised 
  	by $z_\mathrm{w} = 1.0$, $r_\mathrm{wout} = 1.2$~pc, $\zeta = 0.8$ and $r_\mathrm{win} = 0.25$ / $r_\mathrm{win} = 0.5$~pc. The SEDs 
  	corresponding to these models are drawn in 
  	blue for $r_\mathrm{win} = 0.25$~pc, solid for $\phi = 0^\circ$ and dotted for $\phi = 90^\circ$, respectively in black 
  	for $r_\mathrm{win} = 0.5$~pc
  	(dashed-dotted: $\phi = 0^\circ$, dashed: $\phi = 90^\circ$). In addition in the right panel all SEDs, except the two mentioned before, 
  	of the strongly warped models are drawn in green and those of the weakly warped models in red for a rotation angle of $\phi = 0^\circ$.}
  	\label{fig:SED_param_study_w_p0}
\end{figure*}

In order to get an idea of the parameter dependencies of our model, we ran a series of simulations by changing each of the 
parameters defining the warp morphology by a factor of two with respect to our standard model, resulting in 16 simulations
(see table~\ref{tab:parameters}).  
The change of the dust distribution is demonstrated in Fig.~\ref{fig:contour_lines}.
The strongest change in the warp morphology is visible for changes of $r_\mathrm{wout}$. A larger value leads to a weaker 
warp -- or a smaller shift in $z$-direction respectively, e.~g.~compare the 
green dotted line to the blue dashed line in Fig.~\ref{fig:contour_lines_xz_plane}. 
A similar effect can be reached with the parameter $\zeta$, but in opposite direction (compare the green dashed-dotted to the blue dashed line). 
Small differences arise due to
the additional change of the range of the warp and the corresponding smooth continuation beyond $r_\mathrm{wout}$.
Parameter $z_w$ scales the warp function, partly leading to sharp edges towards the inner unwarped region.
$r_\mathrm{win}$
has a minor effect on the overall warp, only defining the width of the inner, unwarped disc (compare the blue dashed with the black, long-dashed
line). Hence, other models with $r_\mathrm{win}=0.5$ are not discussed further.
Overall, the simulations can be split into two categories: (i) strongly warped discs (left panels of Fig.\ref{fig:contour_lines}a and b) 
that extend far from the $xy$-plane and (ii) minor warps which remain close to the flat model 
(right panels of Fig.\ref{fig:contour_lines}a and b).
Concerning the spectral energy distributions under an inclination angle of 45$^\circ$ (Fig.~\ref{fig:SED_param_study_w_p0}, left panel), 
all parameter changes lead to 
minor changes in the SEDs, as for this inclination angle all lines of sight are 
within the dust-free region. Therefore, this also applies to smaller inclination angles and differences are only expected due to 
the inner rim of the unwarped, central disc.
The exception are the models with 
$z_\mathrm{w}=1.0$~pc, $\zeta=0.8$ and $r_\mathrm{wout}=1.2$~pc and $r_\mathrm{win}=0.25$~pc and 0.5~pc which 
extend furthest in $z$-direction (Fig.~\ref{fig:contour_lines}, blue dashed and black long dashed lines). 
In this case, some of the lines of sight are now able to intersect with cold dust, 
leading to partly strong extinction, 
substantially decreasing the flux shortwards of ~50\text{\SI{}{\micro\meter}}.
Hence, the range of SEDs for various $\phi$ angles increases a lot towards lower energy 
output. This can be seen for the spectral energy distributions of the model with 
$z_\mathrm{w}=1.0$, $\zeta=0.8$ and $r_\mathrm{wout}=1.2$ for the line of sight for which 
the highest extinction is reached ($\phi=0^\circ$, blue solid and black dash-dotted line in Fig.~\ref{fig:SED_param_study_w_p0}, left panel).
For an inclination angle of 90$^\circ$, the parameter changes hardly affect the line of sight along the $y$-axis 
(along the unwarped direction, $\phi=90^\circ$). This is shown by the dark grey area in Fig.~\ref{fig:SED_param_study_w_p0}, right panel, 
and is expected because the density
distribution in this direction is least affected by parameter changes. The only notable exception is given by 
the parameter set $z_\mathrm{w}=1.0$, $\zeta=0.8$, $r_\mathrm{win}=0.5$ and $r_\mathrm{wout}=1.2$. 
In this case, the central unwarped disc extends to similar or even greater distances than 
the outer unwarped fraction of the disc (perpendicular to the $y$-axis) in the outer part (see black, long-dashed line in Fig.~\ref{fig:contour_lines}, right panel). 
This slightly boosts the flux at short wavelengths and is able to partly fill up the silicate feature in
absorption (see black, long-dashed line in Fig.~\ref{fig:SED_param_study_w_p0}, right panel). 
For lines of sight along the x-axis ($\phi=0^\circ$, parallel to the direction of the strongest warp), we find a dichotomy:
the mildly warped models all show the silicate feature in absorption only (red lines in Fig.~\ref{fig:SED_param_study_w_p0})
and do not show significant differences for 
the change in $r_\mathrm{win}$. The reason being that there is always some cold dust along the line of sight. 
Only those models exhibiting a strong enough warp such that some $\phi$ angles allow direct lines of sight towards the 
inner, unwarped disc show a larger variety of SEDs. This also leads to a significant difference when changing the inner
warp radius, as $r_\mathrm{win}=0.25$ allows direct views towards hot dust, resulting in the silicate feature to appear in emission.
This parameter study is meant to give a rough impression of the importance of the strength of the warp. As for any
geometrical toy model, there is a large amount of degeneracy that significantly impacts on the resulting SEDs and images
(see discussion in Sect.~\ref{sec:discussion}).

\section{Comparison to the Circinus Galaxy}
\label{sec:circinus_comparison}

In this section we apply our warped standard model\footnote{The discussion will be limited to a comparison of the 
standard model, which has been chosen such to give the best representation of the data.} to the specific case of the Circinus galaxy and compare it with 
observations by~\citet{Tristram_14}. 
Our goal is not an accurate modeling of these observations, but rather to qualitatively assess the applicability and 
limitations of the model at hand.
For the comparison with the Circinus galaxy SED we assume a distance of $4.2\,\text{Mpc}$ 
in accordance with~\citet{Freeman_77}. As already mentioned, \citet{Tristram_14} find that the 
interferometric data in the MIR is best described by a two-component model of the 
brightness distribution. It is composed of a diffuse large scale component, which 
seems to be associated with the ionisation cone and narrow line region (NLR)
and an elongated, elliptical, disc-like component, which includes a surface brightness
concentration that is offset from the centre. 
Here we are only interested in the disc-like component, which -- when combined with 
the maser emission data -- might be explicable by a warped disc.
To this end, we will limit our discussion in the following to a comparison of the 
modelled brightness distribution as well as the modelled SEDs of the point source plus 
disc-like component from \citet{Tristram_14}, which can be separated by comparing to 
certain baseline configurations of the interferometer.
We also discuss the corresponding geometrically flat model in order to examine the differences with respect to the warped case.

\subsection{Warped case}
\begin{figure*}
	\centering
  	\subfloat[]{\includegraphics[width=0.4\textwidth]{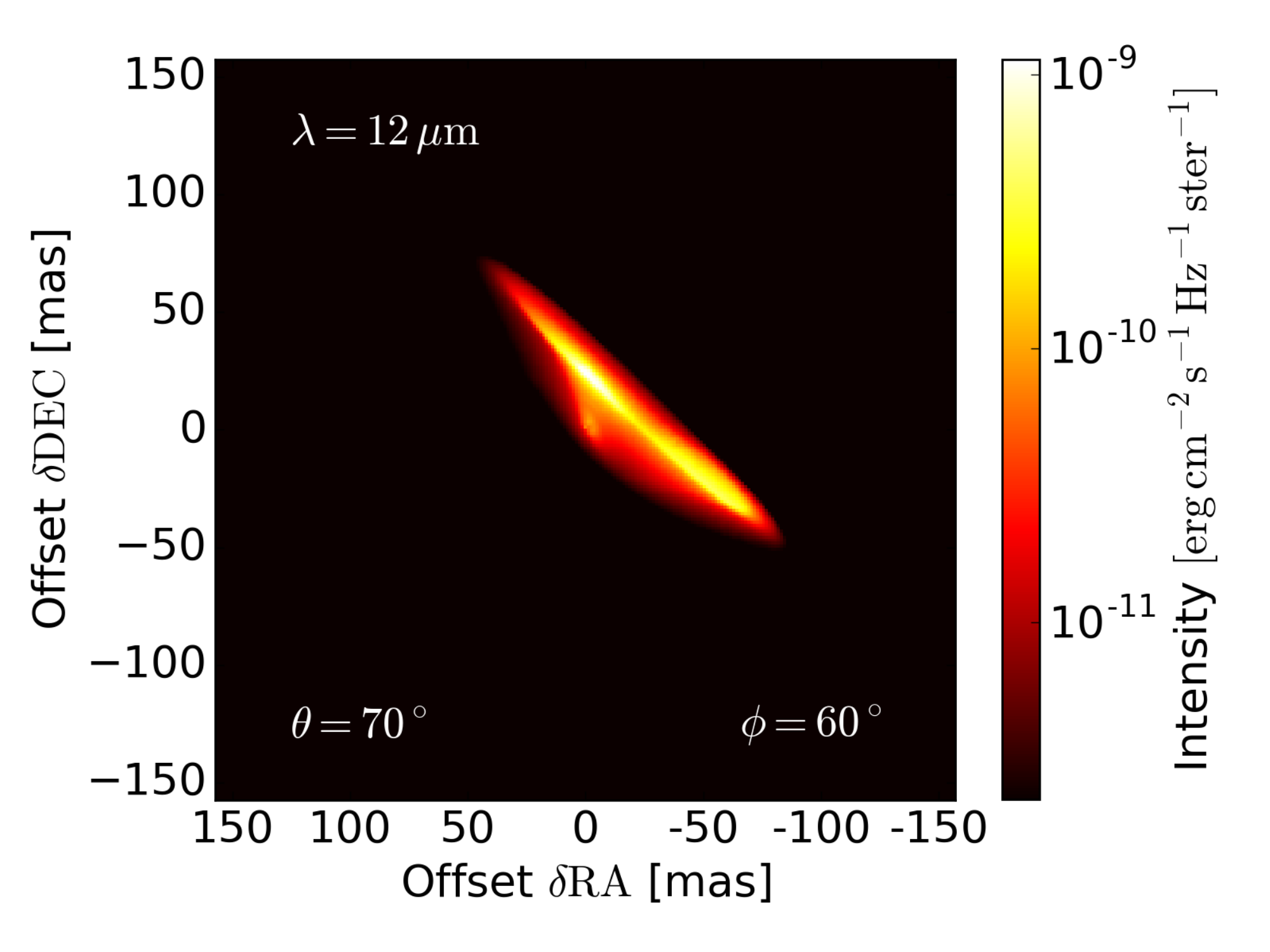}\label{fig:Image_Circinus}}
	\hspace{4pt}
  	\subfloat[]{\includegraphics[width=0.4\textwidth]{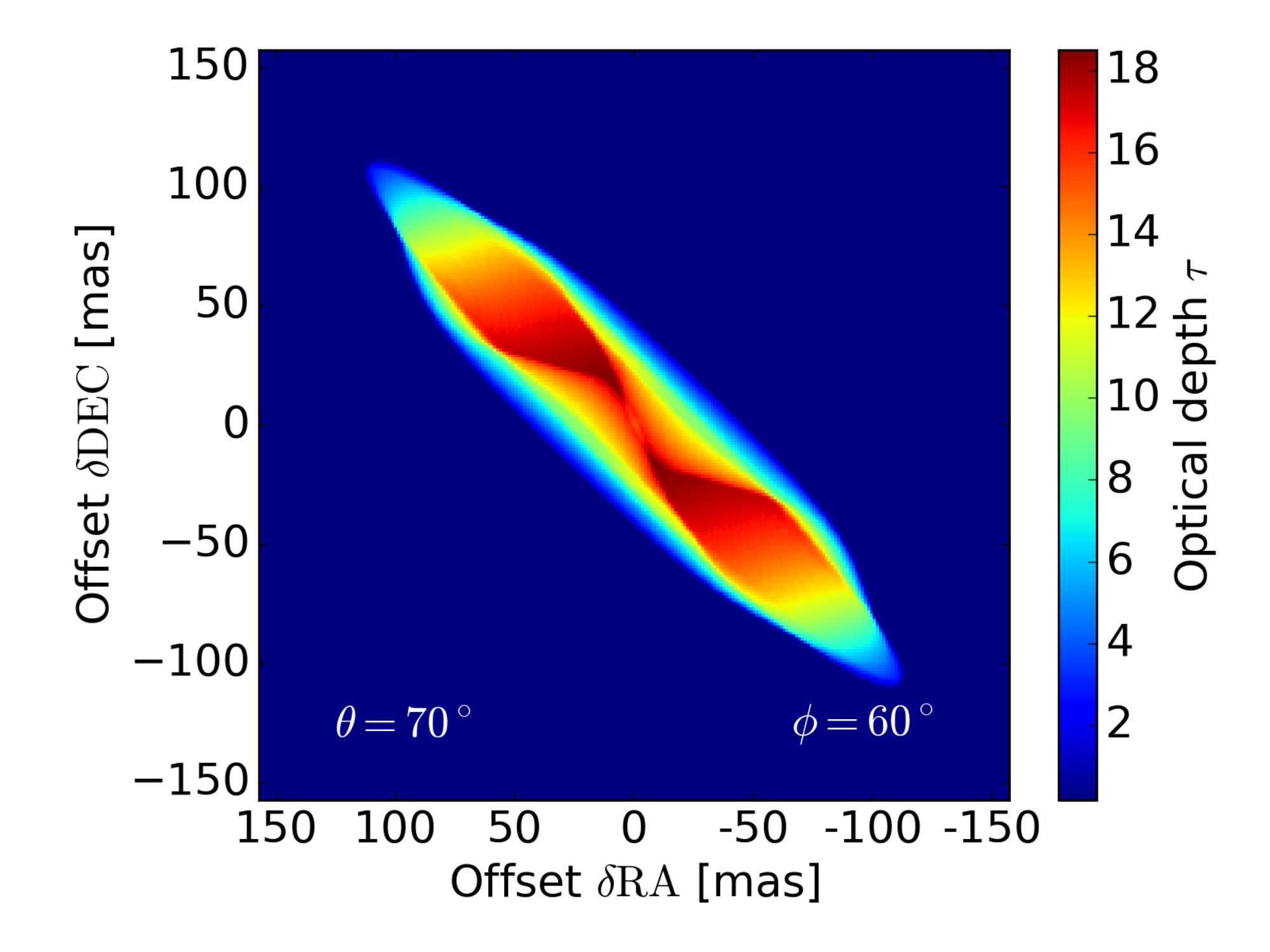}\label{fig:Tau_Circinus}}
  	\hspace{4pt}
  	\subfloat[]{\includegraphics[width=0.5\textwidth]{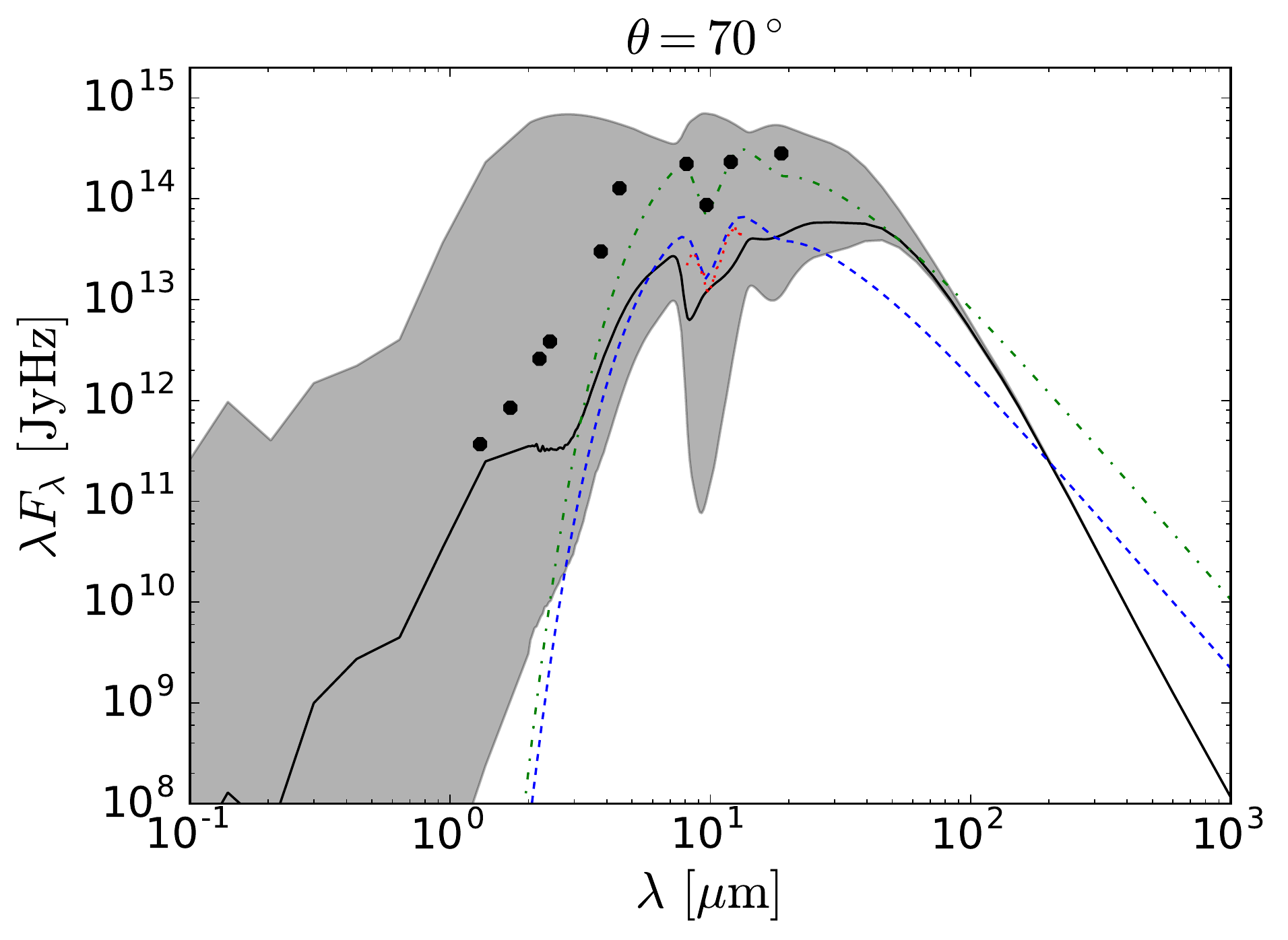}\label{fig:SED_Circinus}}
  	\caption{Overview over the results based on the warped dust distribution (our standard model) which fit best the observations by~\citet{Tristram_14} 
  	(compare to their Fig.~7). In~(a) we show a rotated image at 12\,\text{\SI{}{\micro\meter}} for $\theta = 70^\circ$ 
  	and $\phi = 60^\circ$. In~(b) we plot the optical depth at 12\text{\SI{}{\micro\meter}} for the same orientation. Finally, in~(c) we draw the SED for this 
  	line of sight in solid black.
The observed high-spatial resolution SED compiled by~\citet{Prieto_10} is shown by filled dots. In addition, we show the best-fit 
model to the MIDI data described in \citet{Tristram_14} which consists of two components (disc, large scale component with point source). The disc 
together with the point source is drawn dashed blue while the total model including the large scale component is drawn dashed 
dotted green. In dotted red we show the MIDI correlated flux spectrum of a long baseline concentrating on the disc-like component.
Within the region shaded in grey lie all SEDs with lines of sight described by $\theta = 70^\circ$ and $\phi$ taking all values 
in the range $[0^\circ,180^\circ]$ with a step size of 10 degrees. The SEDs are normalised to a distance of $4.2\,\text{Mpc}$.}
  	\label{fig:Circinus}
\end{figure*}

For our standard model, we find the best comparison to the data for an observer with a line of sight described by an inclination of $\theta = 70^\circ$ and $\phi = 60^\circ$,
 corresponding to an almost grazing line of sight. 
The total dust mass in the model was adjusted
to obtain a good correspondence to the silicate feature depth in the MIDI correlated flux spectrum.
 The inclination angle of $\theta = 70^\circ$ is in agreement with the inclination angle of the galactic disc of the Circinus galaxy \citep[$i=65^\circ$,][]{Freeman_77}. 
 However, \citet{Greenhill_03} estimate a roughly edge-on orientation as the masers form a tight spatial sequence.
 In the leftmost plot of Fig.~\ref{fig:Circinus} 
we show the mid-infrared emission at a wavelength of \SI{12}{\micro\meter} for this observer position. We choose a logarithmic scale for the intensity such that also fainter parts of the disc can be seen. Looking only at the high intensity regions we can observe a three-component structure: 
(i) The warped regions which are directly illuminated by the central source are only visible partially. They hence appear as an elongated {\it disc-like} structure
which is asymmetric and displaced to the lower right with respect to the position of the central source. No direct signs of a warped density distribution is visible in this intensity map. (ii) A brightness concentration within this disc, which is offset to the upper left and (iii) a much fainter {\it blob} component reflecting the innermost part of the directly illuminated warp and is located very close to the heating source. 
Neglecting the latter which only contributes approximately 1-2\% of the total flux, this is in accordance to the disc component together with the point-like concentration of the model by~\citet{Tristram_14}. 
Despite being a crude first model, the flux contribution of the disc ($\approx$ 70-75\%) and point source ($\approx$ 25-30\%) w.~r.~t.~the sum of both is in reasonable agreement with the model presented in \citet{Tristram_14}, who find
values of 70\% and 30\% respectively.
Mind that the MIDI observations described there do not allow for accurate positioning of the components. Especially the point source component should rather be regarded as a
hint towards an asymmetric distribution in the disc-like component. 
In the context of our warped disc model, we hence interpret the asymmetric disc in the visibility data as evidence for a warped dust distribution.
Another interesting prediction of our model is the offset of the disc-like component in north-west direction by approximately 20\,mas with respect to the galactic nucleus.

Maser emission from our model is expected along the lines of sight with the largest optical depth. The distribution of the optical depth at 12\text{\SI{}{\micro\meter}} for the same inclination and rotation angle is shown in the middle panel of Fig.~\ref{fig:Circinus} 
and clearly shows a warped disc morphology. 
Therefore, our choice of the dust distribution and line of sight seems to be able to simultaneously account for the MIR interferometric results, as well as a warp similar to 
the maser disc geometry. Finally, in the right panel of Fig.~\ref{fig:Circinus}, we show the corresponding SED as the black solid line, together with the model 
of \citet{Tristram_14} and high spatial resolution observations compiled by \citet{Prieto_10}. 
The VLTI/MIDI observations allow us to disentangle the nuclear mid-IR emission. At a position angle of roughly $125^\circ$ and a projected baseline length 
of roughly 60~m, the extended component is essentially resolved and the disc component is essentially unresolved. The observed correlated flux at this position in the (u,v) plane can 
therefore be considered as emission only from the disc plus point source component (red dotted line, \citealp{Tristram_14}),
which is in reasonable agreement with our simulation.
The overall scale of our modelled SED is lower than the single-dish observations 
(filled dots), which is understandable given the small covering factor of our disc and 80\% of the total MIR flux coming from the extended 
emission component in \citet{Tristram_14}.

\begin{figure*}
	\centering
  	\includegraphics[width=\textwidth]{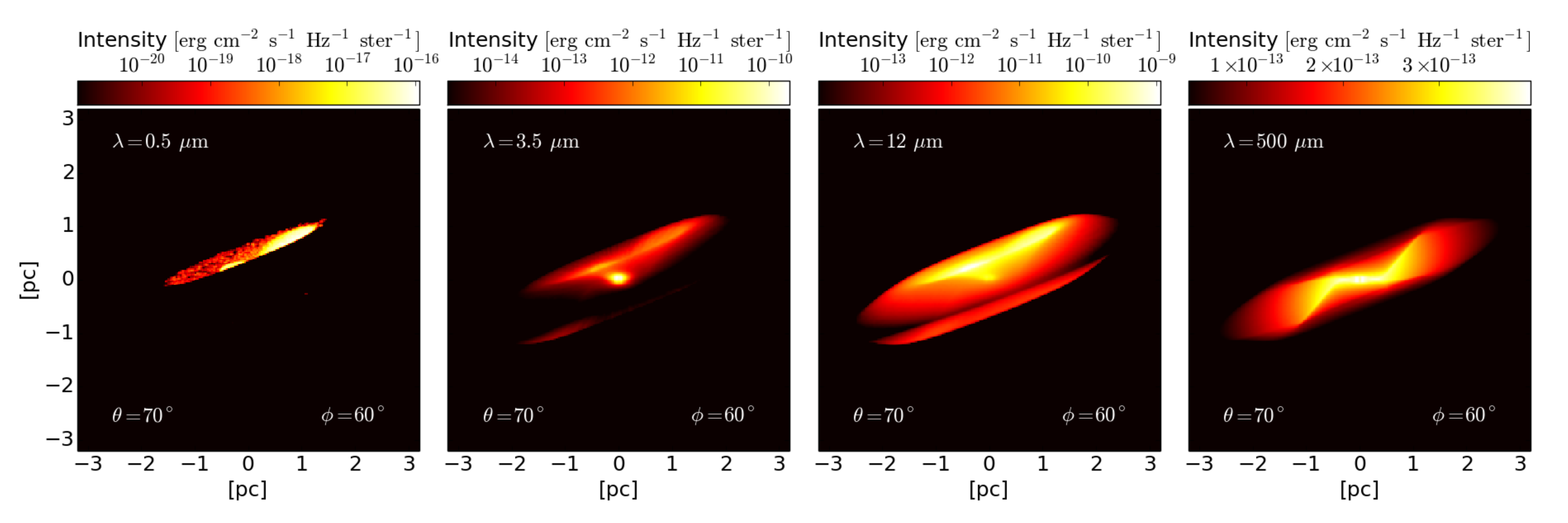}
  	\caption{Images of our standard model for different wavelengths ($\lambda \in \left\{ 0.5, 3.5, 12, 500 \right\}\,\text{\SI{}{\micro\meter}}$) and a line of sight as in Fig.~\ref{fig:Circinus} ($\theta = 70^\circ$, $\phi = 60^\circ$). For the \SI{0.5}{\micro\meter} image we use a larger number of photons for the scattering Monte Carlo run of $n_\mathrm{phot\_scat} = 5\times 10^{7}$ due to the large number of scattering events for high energy photons. Note that the image at \SI{12}{\micro\meter} is the same as in Fig.~\ref{fig:Circinus} but with a larger intensity range and for the image at \SI{500}{\micro\meter} we used a linear intensity scale. All parts with a lower intensity than the minimum shown in the legend are plotted in black. Large morphological differences are found in the appearance of the warped disc in the visible to far-infrared (FIR) wavelength range.}
  	\label{fig:Circinus_wavelength}
\end{figure*}

To better characterise the model, we show images for different wavelengths and the same line of sight ($\theta = 70^\circ$, $\phi = 60^\circ$) in Fig.~\ref{fig:Circinus_wavelength}. 
In the leftmost plot, where we show an image in the visible range of the spectrum ($0.5\,\text{\SI{}{\micro\meter}}$), 
we only see a disc-like structure which seems to be broken up into two components. 
The L-band ($3.5\,\text{\SI{}{\micro\meter}}$) image in the second panel looks more similar to the \SI{12}{\micro\meter} image discussed before (and shown in the third panel). However, tracing the emission from hotter dust, the central region is more pronounced in the $3.5\,\text{\SI{}{\micro\meter}}$ image compared to longer wavelengths. Only a small fraction of the directly illuminated warp of the disc reaches temperatures high enough to be visible in the L-band. 
This suggests that images reconstructed from upcoming MATISSE \citep{Lopez_14} L-band observations might be dominated by a point source component.
The rightmost panel corresponds to an image at a FIR wavelength of $500\,\text{\SI{}{\micro\meter}}$. Interestingly, the structure changes significantly and our model 
predicts that the warp should be recoverable with Atacama large millimeter array (ALMA) observations, which in the most extended configuration of 16~km will reach a spatial resolution of roughly 8\,mas.  
However, one should keep in mind that our model does not include the large scale component along the ionisation cone, which might contribute significantly.

\begin{figure}
	\centering	
 	\includegraphics[width=200px]{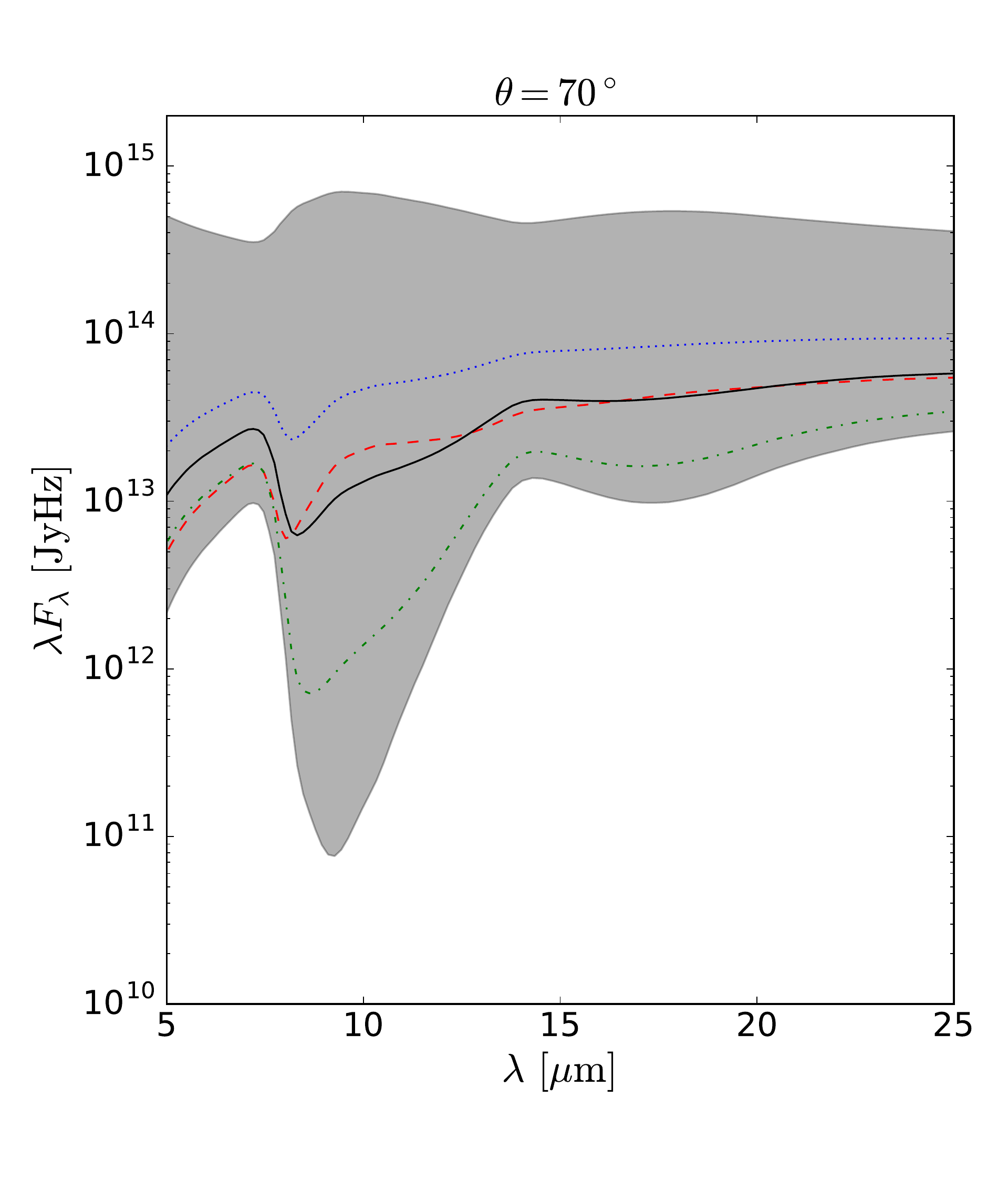}
 	\caption{Close on view on the part of the SEDs of our standard model corresponding to the wavelengths associated with the silicate features. We draw in solid black the line of sight corresponding to our best fit model ($\theta = 70^\circ$, $\phi = 60^\circ$) together with two lines of sight corresponding to $\phi = 50^\circ$ (dotted-dashed green) and $\phi = 70^\circ$ (dotted blue) for the same inclination. Furthermore, we show the SED for an observer characterised by $\phi = 0^\circ$ and $\theta = 70^\circ$ (dashed red). The region between the minimal and maximal values obtained for the SEDs in the range described by the angles $\theta = 70^\circ$ and $\phi \in [0^\circ,180^\circ]$ with a step size of 10 degrees is shaded in grey. We normalised the SEDs to a distance of $4.2\,\text{Mpc}$.
Various silicate feature shapes can be found depending on the orientation.}
 	\label{fig:Circinus_feature}
\end{figure}

To emphasise that our model shows a large variety of SEDs and feature shapes even for small changes in the line of sight, we show in Fig.~\ref{fig:Circinus_feature} a close on view on the wavelength range containing the characteristic silicate features for observers at a location around the best fit angles. As we see in this plot the shape of the feature at \text{\SI{9.7}{\micro\meter}} changes noticeably even for small variations of the $\phi$ angle (the same is true for changes in the $\theta$ angle). 
We see that the features for $\phi = 50^\circ$ (dotted-dashed green) and $\phi = 70^\circ$ (dotted blue) show both a different shape and a different strength as compared to the best fit angles. 
This strong dependence on the line of sight is associated with the complex geometry of our warped disc for viewing angles close to grazing lines of sight.
Only a small change in viewing direction can significantly decrease the column density or increase the amount of directly illuminated dust observed. This has the effect that an absorption feature is 
filled up from the long wavelength end (by adding emission components), resulting in a change of slope of the feature towards longer wavelengths. 
The latter effect is independent of the geometrical distribution, but the shift of the minimum (or maximum in the case of an emission feature) happens when the optical
depth of the dust in-between the source and the observer changes. Interestingly, for our model geometry, this happens for grazing lines of sight at inclinations for which we also reach the 
best agreement with the observations of the Circinus galaxy. This is not easily the case for a doughnut-like structure.
A similar difference of the silicate feature 
shape w.~r.~t.~the absorption profile for a typical galactic dust mixture has been seen in the nearby Seyfert~2 galaxy 
NGC~1068 \citep{Raban_09}. 
They found that an absorption profile with a shallower slope towards longer wavelengths compared to the standard galactic one is needed in order to be able to explain their MIDI observations 
(see Fig.~5 in \citealp{Raban_09}). 
Our study shows that
a significant change can already be obtained in a flat disc geometry when slightly changing the observing direction around grazing lines of sight (see also Sect.~\ref{sec:flat_case}). 
This geometrical effect could provide an alternative (and probably more natural) explanation for the seemingly unusual shape of the silicate feature found in NGC~1068, which harbours a 
warped maser disc as well.
In contrast, the observed silicate feature in the Circinus galaxy can be well explained by a standard galactic dust mixture and is at odd with our best-fit model. 
Viewing angles with more dust along the line of sight result in a better match, e.~g.~viewing along $\theta=\phi=90^\circ$ (green dot-dashed line in Fig.~\ref{fig:SED_incl}).
The same effect could also be obtained by adding an absorption screen, which 
could be provided by the missing large scale component.  
In Fig.~\ref{fig:Circinus_feature} we also show the SED for an observer characterised by $\theta = 70^\circ$ and $\phi = 0^\circ$. Even though the depth of 
the feature at \SI{9.7}{\micro\meter} is of comparable size to the best fit model it shows a different shape, 
which is specific to our choice of warped disc geometry. 
From a flat model we would expect that the change in shape is always also accompanied by a change in the depth of the feature. The reason being that we have only one free parameter 
characterising the line of sight. Therefore we always change the column density along the line of sight and thus the depth of the feature by changing this parameter. 

%

%

\subsection{Flat case}
\label{sec:flat_case}

\begin{figure*}
	\centering
  	\subfloat[]{\includegraphics[width=0.3\textwidth]{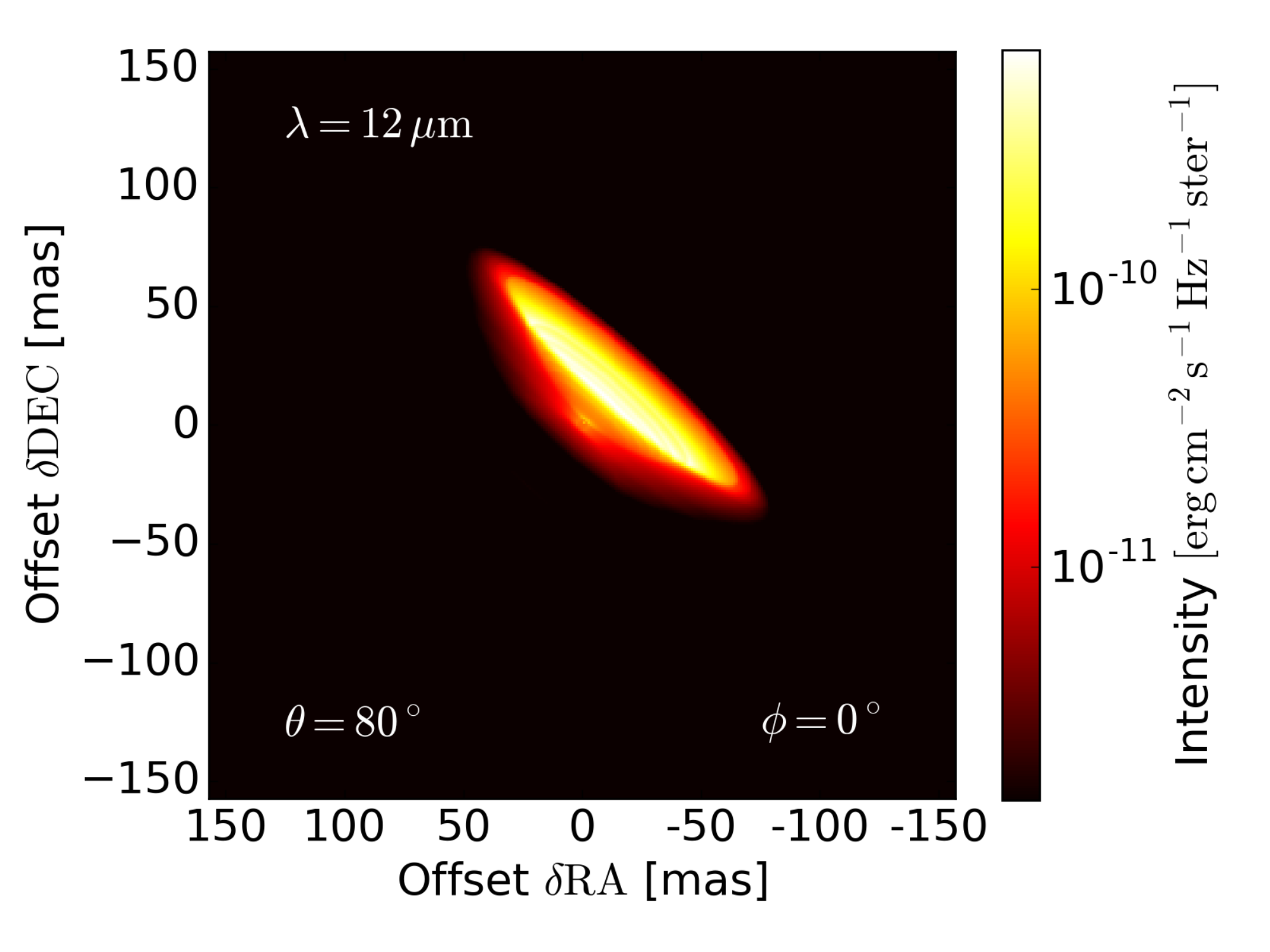}\label{fig:Image_Circinus_flat}}
	\hspace{4pt}
  	\subfloat[]{\includegraphics[width=0.3\textwidth]{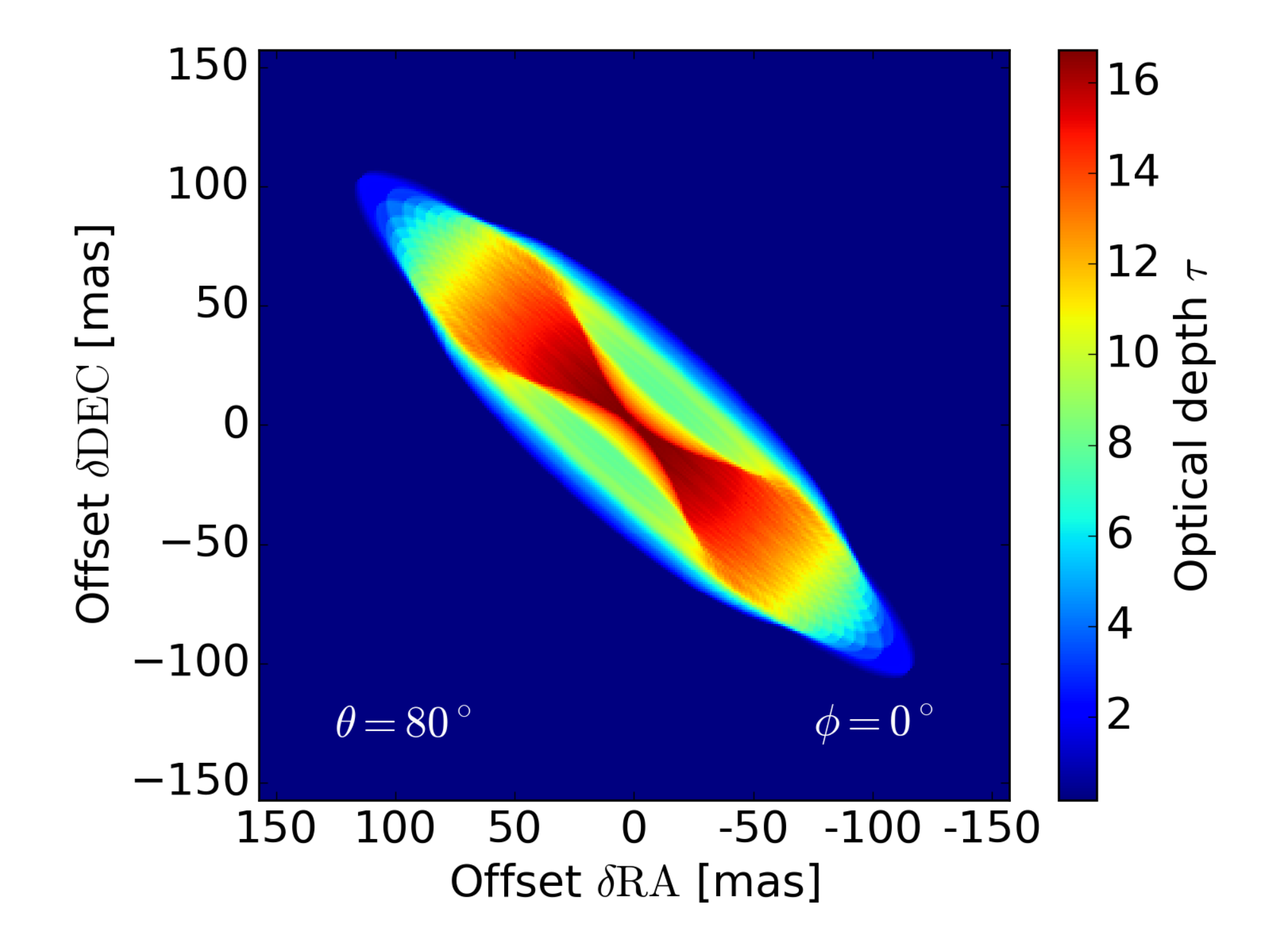}\label{fig:Tau_Circinus_flat}}
  	\hspace{4pt}
  	\subfloat[]{\includegraphics[width=0.3\textwidth]{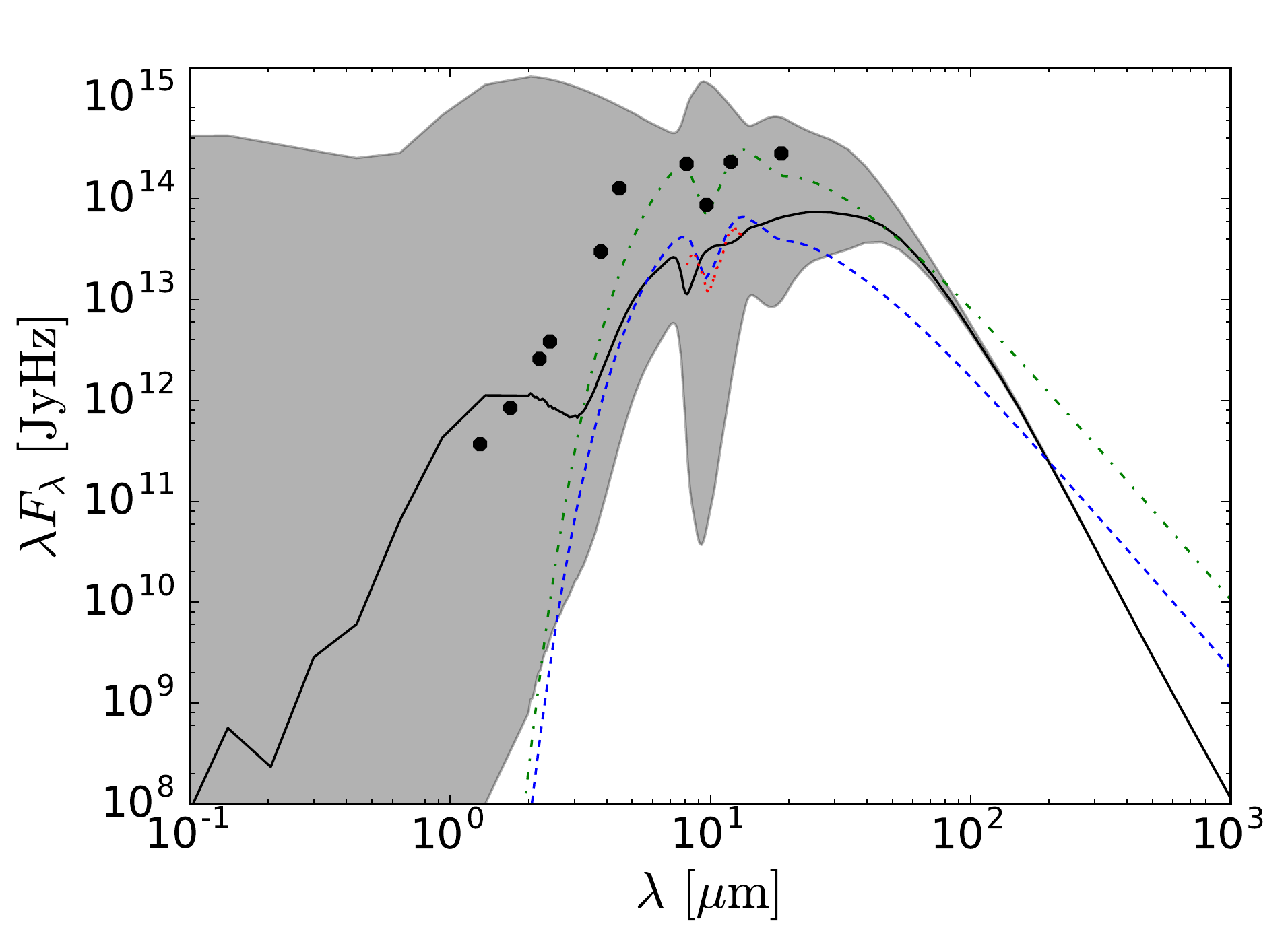}\label{fig:SED_Circinus_flat}}
  	\caption{The best comparison of the flat model (corresponding to our standard model) to the Circinus observations is reached for a line of sight with 
  	$\theta=80^\circ$ and $\phi=0^\circ$. Panel (a) shows an image at 
\SI{12}{\micro\meter}, which has been rotated to fit the appearance on the sky (compare to Fig.~7 in \citealp{Tristram_14}). 
In~(b) we plot the optical depth at 12\text{\SI{}{\micro\meter}} for the same orientation. Panel (c) shows the SED corresponding to $\theta = 80^\circ$ in 
solid black and we overlay the same observations as in Fig.~\ref{fig:Circinus}.   	
 Within the region shaded in grey lie all SEDs with lines of sight described by $\theta \in [0^\circ,90^\circ]$ with a step 
 size of 10 degrees. The SEDs are normalised to a distance of $4.2\,\text{Mpc}$.}
  	\label{fig:Circinus_flat}
\end{figure*}

In Fig.~\ref{fig:Circinus_flat} we show the same type of plots as in Fig.~\ref{fig:Circinus} but for the case of an unwarped disc with otherwise unchanged parameters.
 The line of sight which fits best the model of~\citet{Tristram_14} is in the flat case described by an inclination angle of $\theta = 80^\circ$. We plot the mid-infrared emission at a 
 wavelength of \SI{12}{\micro\meter} in the left panel of Fig.~\ref{fig:Image_Circinus_flat}. The basic structures we observe are a dominating elongated disc-like component and a 
 much fainter point source. Due to symmetry and in contrast to the warped case, the disc does not possess any sub-structure or asymmetry that could be associated to 
 the point source component of the MIDI models. 
 Clearly we see in the middle panel of Fig.~\ref{fig:Tau_Circinus_flat} that the optical depth distribution has a perfectly flat major axis and -- of course -- no traces of a warp in the maser emission are expected to be seen. Nevertheless, the line of sight can be chosen such that we can find accordance in the SED with the model by~\citet{Tristram_14} when restricting to the disc and point-like component, as shown in Fig.~\ref{fig:SED_Circinus_flat}. 
If we compare the shape of the silicate features to Fig.~\ref{fig:Circinus_feature} we see that in the flat case the feature changes its shape as well. However, as we have only one free parameter describing the line of sight in the flat case we can not find two lines of sight with comparable feature depth but completely different shape in the SED around the feature.

\section{Discussion}
\label{sec:discussion}

The results we present resemble available observations of warped discs in many respects, e.~g.~when looking at the 
SED (restricted to the small-scale disc and point source component) and the geometrical implications as derived from MIDI observations. 


For our calculations we assumed an isotropically emitting central source. However, a geometrically thin, optically thick accretion disc would lead to an anisotropic emission, where the flux depends on the polar angle $\theta $ as $F \propto \cos \theta ( 1+2\,\cos \theta )$~\citep[e.~g.~][]{Netzer_87}. This would imply a stronger emission in direction of the disc axis, while the emission in direction of the plane would be reduced \citep{Schartmann_05, Manske_98}. However, the orientation of the inner accretion disc with respect to our dust structure is unclear, especially for asymmetric discs as in the case of a warp. To this end, an isotropically emitting source seems to be most appropriate for this first study of warped disc geometries.
Given the large inclination angles used for our comparison, we expect only a minor change of the short wavelength end of the SED \citep[Fig.~12,][]{Schartmann_05}.

Another point for discussion is the large degeneracy we introduce with our dust distribution with a total of 8 parameters. 
This means that we can achieve a large variety of different disc geometries and emission characteristics, 
sometimes producing the same overall predictions even with quite different parameters (see Sect.~\ref{sec:par_study}). 
In addition, by choosing the line of sight appropriately, we observe large differences in the observed spectra. 
In this first approach towards studying warped discs as alternatives or amendments of classical torus models, we want to 
concentrate on general 
properties and whether such discs might be able to describe some of the peculiar properties of the Circinus galaxy.
However, there may be a better 
combination of parameters and line of sight to explain the observations. 
Given the promising results of the phenomenological model presented here, a more detailed investigation of possible 
structures of warped discs and possible physical mechanisms that 
might cause such warps would be worth investigating further.

In addition, due to the constant dust density the calculations are simpler and we achieve converged results for a lower spatial resolution of the 
coordinate grid. Nevertheless, in a more physical approach there would likely be a density gradient along the radial 
direction \citep[e.~g.~][]{Schartmann_05}. However, as the break of axisymmetry in warped discs requires 3D simulations, we 
currently cannot achieve sufficiently converged results for the case of a steep density gradient. Therefore, one would either need a 
larger overall resolution or refine the grid in the inner regions of the torus where the dust is heated to the highest temperatures.

%

The simulations presented here are a first study of the influence of warped disc geometries on observable quantities.
We neglect several things: (i) There are many indications that tori need to be clumpy. 
For the sake of simplicity, we concentrate on smooth dust distributions within warped geometric discs in this study. Changing this into a clumpy distribution will affect the shape of the silicate feature \citep{Nenkova_02,Hoenig_06,Schartmann_08} and the SEDs in general. 
(ii) Taking grain dependent sublimation into account has been shown to have a significant effect, especially for the short wavelength end of the IR bump \citep{Schartmann_05}. 
We defer these modifications to future work. This might include the combination of the simulations of this article 
with the larger scale structure found in a recent radiation hydrodynamical simulation of the core of the Circinus galaxy \citep{Wada_16}.

\section{Summary and Conclusions}
\label{sec:conclusions}

In this article, dust continuum radiative transfer calculations have been presented to characterise the observable signatures of dusty warped discs in the infrared. 
Our main results can be summarised as follows:

\begin{enumerate}
\item Depending on the orientation of the warp with respect to the line of sight, we find strong variations of the brightness distributions. 
The geometry of the underlying warped density distribution can only be recovered at long wavelengths, whereas short wavelengths typically show
thin discs and point sources in the brightness distribution.
\item Applying the models to the best-studied case of the Circinus galaxy,
      we find that a dust density distribution in form of a warped disc 
      is able to account for the inner disc-like emission component revealed by MIDI observations.  
      The warped disc geometry results in an asymmetric appearance (with respect to the central black hole) of the disc-like component for an inclination angle of around 70 degrees.
      The best-fit inclination needs to be close to a grazing viewing position and is in agreement with what is inferred for the galactic disc of the Circinus galaxy by other 
      measurements as well. Also the point source component  
      arises naturally on the visible, directly illuminated side of the warped disc structure and is offset from the centre point. This is in line with the original interpretation of 
      the point source being part of an asymmetric disc, as its location could not be very well determined.
      The optical depth / column density distribution at this orientation clearly shows a 
      warped disc geometry (Fig.~\ref{fig:Tau_Circinus}). This represents the highest probability for maser emission in our model, which was chosen to resemble this observed warp geometry. 
      At the same viewing angle, the dust emission maps in the MIR (Fig.~\ref{fig:Image_Circinus}) resemble a very thin, extended structure. No signs 
      of the underlying warped disc geometry can be found in the emission in this wavelength range.
\item For the same orientation of the disc, we find qualitative agreement of the simulated SED with the MIDI correlated flux spectrum corresponding to a long baseline, which 
      is mostly sensitive to the disc emission, excluding the large scale component mentioned before.      
\item The disc orientation has significant influence on the spectral shape of the silicate feature for grazing lines of sight, resulting in a change of slope towards the long 
      wavelength end of the feature. This is reminiscent of MIDI correlated flux spectra of the Seyfert~2 galaxy NGC~1068, which harbours a (slightly) warped disc component on 
      parsec scale as well. Thereby, it provides an alternative
      explanation to the claim of a non-standard dust composition in this nucleus. 
\end{enumerate}

Overall the simulations indicate that a central, parsec-scale, warped, dusty disc provides remarkable features identifiable in some high-resolution observations of 
nearby Seyfert galaxies, which make it an attractive amendment to classically assumed AGN torus or newly established 
outflow morphologies.

\section*{Acknowledgements}

We thank C.~P.~Dullemond for making {\sc RADMC-3D}\footnote{\url{http://www.ita.uni-heidelberg.de/~dullemond/software/radmc-3d/}} publicly 
available and as user friendly as it is.




\bibliographystyle{mnras}
\bibliography{literature}





\bsp	
\label{lastpage}
\end{document}